\documentclass[twocolumn,tighten,twocolappendix]{aastex631}
\usepackage{CJK}
\usepackage{amsbsy}
\usepackage{multirow}
\usepackage{bm}
\usepackage{ulem}
\usepackage{graphicx}     % 插入图片
\usepackage{chngcntr}     % 控制计数器

\usepackage{epsfig}
\usepackage{color}
\usepackage{graphicx}
\usepackage{longtable}
\usepackage{hyperref}
\usepackage{amsmath}

\def\kms{{\,\rm km\,s^{-1}}}

\def\e{\,{\boldsymbol{e_1}}}

\def\dzmax{\,{\rm \Delta Z_{Max}}}

\received{XXX}
\revised{YYY}
\accepted{ZZZ}
\published{HHH}

\counterwithout{figure}{section}
\begin{document}
\begin{CJK*}{UTF8}{gbsn}

%%%%%%%%%%%%%%%%%%%%%%%%%%%%%%%%%%%%%%%%%%%%%%%%%%%%%
%%%%%%%%%%%%%%%%%%%
%%%    Paper  Title 
%%%%%%%%%%%%%%%%%%%
%\title{Cosmic galaxy cluster spin}
\title{The Cosmic Dance: Observational Detection of Coherent Spin in Galaxy Clusters}

%%%%%%%%%%%%%%%%%%%%%%%%%%%%%%%%%%%%%%%%%%%%%%%%%%%%%
%%%%%%%%%%%%%%%%%%%
%%%    Author List
%%%%%%%%%%%%%%%%%%%

\correspondingauthor{Peng Wang}
\email{pwang@shao.ac.cn}

\author[0009-0001-7527-4116]{Xiao-xiao Tang (唐潇潇)}
\affil{Shanghai Astronomical Observatory, Chinese Academy of Sciences, Nandan Road 80, Shanghai 200030, People's Republic of China.}
\affil{University of Chinese Academy of Sciences, Beijing 100049, People's Republic of China.}

\author[0000-0003-2504-3835]{Peng Wang* (王鹏)}
\affil{Shanghai Astronomical Observatory, Chinese Academy of Sciences, Nandan Road 80, Shanghai 200030, People's Republic of China.}

\author[0000-0002-2204-6558]{Yu Rong (容昱)}
\affiliation{Department of Astronomy, University of Science and Technology of China, Hefei, Anhui 230026, People's Republic of China.}
\affiliation{School of Astronomy and Space Sciences, University of Science and Technology of China, Hefei 230026, Anhui, People's Republic of China.}

\author{Weiguang cui (崔伟广)}
\affiliation{Departamento de F\'isica Te\'orica, M-8, Universidad Aut\'onoma de Madrid, Cantoblanco, E-28049, Madrid, Spain.}
\affiliation{Centro de Investigaci\'on Avanzada en F\'isica Fundamental (CIAFF), Universidad Aut\'onoma de Madrid, Cantoblanco, E-28049 Madrid, Spain.}
\affiliation{stitute for Astronomy, University of Edinburgh, Royal Observatory, Edinburgh EH9 3HJ, United Kingdom.}

\author[0009-0005-9342-9125]{Min Bao (鲍敏)}
\affil{School of Astronomy and Space Science, Nanjing University, Nanjing 210023, China}
\affil{Key Laboratory of Modern Astronomy and Astrophysics (Nanjing University), Ministry of Education, Nanjing 210023, China}

%%%%%%%%%%%%%%%%%%%%%%%%%%%%%%%%%%%%%%%%%%%%%%%%%%%%%

%%%%%%%%%%%%%%%%%%%%%%%%%%%%%%%%%%%%%%%%%%%%%%%%%%%%%

\begin{abstract}

The spin of galaxy clusters encodes key information about their formation, dynamics, and the influence of large-scale structure. However, whether clusters possess statistically significant spin and how to measure it observationally remain open questions. 
Here, we present the first observational, statistical detection of coherent spin in galaxy clusters, using two samples of  2,170 and 1,329 systems with $M > 10^{14}\,M_\odot$, selected from two publicly available group catalogs (\citet{2017A&A...602A.100T} and \citet{2012ApJ...752...41Y}) constructed with two different algorithms and but both based primarily on SDSS galaxies.
Cluster spin is quantified by identifying the orientation in the projected plane that maximizes the redshift difference ($\Delta Z_{\rm max}$) between member galaxies in two regions divided by a trial axis. We find compelling statistical evidence for coherent rotation, as the observed $\Delta Z_{\rm max}$ distribution departs markedly from the randomized controls, exhibiting pronounced deviations near $380\,\mathrm{km\,s^{-1}}$. Stacked visualizations confirm the spatial segregation of redshifted and blueshifted galaxies across the rotation axis. The radial profile of the rotational velocity indicates that it increases as a function of radius. The cluster rotation speed increases with mass, from $\sim360~\mathrm{km\,s}^{-1}$ at $10^{14} M_\odot$ to $\sim693~\mathrm{km\,s}^{-1}$ at $10^{15} M_\odot$. Additionally, cluster spin tends to align parallel with the central galaxy spin and perpendicular to the nearest cosmic filament, particularly in richer systems. These results reveal significant coherent spin in galaxy clusters, shaped by both internal dynamics and large-scale structure.

\end{abstract}
%%%%%%%%%%%%%%%%%%%%%%%%%%%%%%%%%%%%%%%%%%%%%%%%%%%%%

%%%%%%%%%%%%%%%%%%%%%%%%%%%%%%%%%%%%%%%%%%%%%%%%%%%%%

%%%%%%%%%%%%%%%%%%%%%%%%%%%%%%%%%%%%%%%%%%%%%%%%%%%%%

\keywords{
    \href{http://astrothesaurus.org/uat/584}{Galaxy clusters (584)};
    \href{http://astrothesaurus.org/uat/902}{Large-scale structure of the universe (902)};
    \href{http://astrothesaurus.org/uat/1882}{Astrostatistics (1882)}}

%%%%%%%%%%%%%%%%%%%%%%%%%%%%%%%%%%%%%%%%%%%%%%%%%%%%%

%%%%%%%%%%%%%%%%%%%%%%%%%%%%%%%%%%%%%%%%%%%%%%%%%%%%%
%%%%%%%%%%
%   Introduction
%%%%%%%%%%
\section{Introduction} 
\label{sec:intro}

The hierarchical formation of structure in a $\Lambda$CDM cosmology proceeds through the anisotropic gravitational collapse of primordial density perturbations, evolving via three distinct dynamical phases. Initially, monodimensional collapse along the first principal axis of the tidal tensor forms \textit{Zel'dovich pancakes}---quasi-2D sheet-like structures---in regions where only one eigenvalue is positive \citep[e.g.,][see references within for more details]{1970A&A.....5...84Z,2014MNRAS.441.2923C}. Secondary collapse along the second axis transforms these sheets into filaments, as mass flows become increasingly anisotropic in regions with two positive tidal eigenvalues. Finally, terminal three-dimensional collapse at filament intersections produces virialized galaxy clusters ($M_{\mathrm{vir}} > 10^{14}M_\odot$), corresponding to density maxima where all three eigenvalues are positive \citep{2018MNRAS.473.1195L}.

These galaxy clusters trace the topology of the cosmic web and typically emerge at the intersection of filaments and walls---zones of matter convergence characterized by high galaxy surface densities ($\Sigma_{\mathrm{gal}} > 10^3\, \mathrm{Mpc}^{-2}$) and extreme accretion rates ($\dot{M} > 1000\, M_\odot\, \mathrm{yr}^{-1}$) \citep{2008LNP...740..335V}. As the most massive virialized structures in the Universe, clusters host galaxy populations ranging from several dozen to thousands, and provide powerful laboratories for probing a wide range of astrophysical and cosmological phenomena---from dark matter (via weak lensing; \citealt{2020A&ARv..28....7U}) and baryonic feedback (through X-ray and radio observations; \citealt{2012gcgc.conf...37M}), to fundamental cosmological parameters such as $\sigma_8$ and $\Omega_m$ \citep{2016A&A...594A..13P}. Cosmological simulations such as Millennium \citep{2005Natur.435..629S} and Illustris \citep{2019ComAC...6....2N} and IllustrisTNG \citep{2018MNRAS.475..648P}  have further confirmed this picture, illustrating how Gaussian initial conditions give rise to the filamentary web of clusters and voids observed today \citep{2018MNRAS.475..676S}.

Among the various dynamical properties of galaxy clusters, \textbf{angular momentum (or ``spin'')} provides a unique window into the anisotropic nature of structure formation. According to \textit{tidal torque theory} (TTT), halo spin arises from torques exerted by the surrounding matter distribution during early structure growth \citep{1969ApJ...155..393P, 1984ApJ...286...38W}. This process links a halo's angular momentum directly to its large-scale environment, particularly the geometry of nearby filaments and walls \citep{2024arXiv240716489K}. While such correlations have been robustly studied in simulations \citep{2012MNRAS.427.3320C, 2015MNRAS.446.2744L, 2018MNRAS.473.1562W}, observational evidence remains limited—especially in the high-mass, nonlinear regime of galaxy clusters. Nevertheless, the mechanisms by which angular momentum is transferred from large-scale structures to the scale of galaxy clusters, and subsequently to individual galaxies within clusters, remain open questions.

Consequently, a pivotal aspect of galaxy clusters is their angular momentum properties. Galaxy clusters can acquire spin through anisotropic infall, primordial torques, or recent off-axis mergers and interactions with neighboring structures. Observational attempts to detect such spins have focused on global rotation or velocity gradient diagnostics. \citet{1992AJ....104.2078O} conducted a detailed study of Abell 2107, finding evidence for global rotation with 98\% confidence. Subsequent works have confirmed this rotational feature through various methods \citep{1983A&A...124L..13M,2005MNRAS.359.1491K}. Broader surveys have revealed significant velocity gradients in subsets of clusters---\citet{1996MNRAS.279..349D} found 13 rotating clusters among 72, while \citet{2004MNRAS.352..605B} reported at least three with possible signs of rotation or shear, indicating velocity gradients or differential motions rather than pure coherent rotation. Analyzing 899 clusters, \citet{2007ApJ...662..236H} found that 6 out of 56 well-sampled systems show rotation under equilibrium conditions. More recently, \citet{2017MNRAS.465.2616M} introduced a novel method to identify both the axis and amplitude of cluster rotation, revealing that $\sim$23\% of clusters exhibit rotation under stringent criteria. Extending this work, \citet{2019MNRAS.490.5017B} employed a ``perspective rotation'' technique to analyze merging systems, finding enhanced core rotation due to red galaxy subpopulations.

Simulations have also provided valuable insights into cluster spin. \citet{2017MNRAS.465.2584B, 2018MNRAS.479.4028B, 2019JPhCS1226a2003B} used MUSIC simulations and kinetic Sunyaev-Zel'dovich (kSZ) maps to detect rotational signals averaging 23\% of the total signal. The MACSIS project \citep{2023MNRAS.524.2262A} further revealed that dark matter, galaxy, and gas spins are often misaligned, with the rotational kSZ signal peaking in massive, unrelaxed clusters. Cosmological implications have also been explored: \citet{2002A&A...396..419C} examined how rotation impacts the CMB temperature and polarization, while \citet{2010ApJ...716L.205H} used line-of-sight velocity mapping to detect tangential motions and cluster kinematics.

Despite these efforts, most prior studies focus on individual or small sets of clusters, lacking a unified statistical framework. In this work, we address this gap by systematically quantifying spin features across a large sample of galaxy clusters using redshift information from the SDSS spectroscopic catalog. Our analysis explores rotation signatures and their dependence on cluster morphology, dynamical state, and environment, providing one of the first statistical assessments of cluster spin in a large observational dataset.

The structure of this letter is as follows. Section~\ref{sec:method} outlines the two galaxy group catalogs and the selection of member galaxies. Section~\ref{sec:result} presents the statistical detection of the cluster spin signal, as well as the connections between cluster spin, central galaxy spin, and the orientation of large-scale filaments. Sections~\ref{sec:sum} and~\ref{sec:dis} provide a summary and discussion of our results. 

Throughout this paper, we adopt the Planck cosmological parameters \citep{2016A&A...594A..13P} with a Hubble constant of $H_0 = 67.8\,\mathrm{km\,s^{-1}\,Mpc^{-1}}$, a matter density of $\Omega_{\rm m} = 0.308$, and a dark energy density of $\Omega_{\Lambda} = 0.692$.

%%%%%%%%%%%%%%%%%%%%%%%%%%%%%%%%%%%%%%%%%%%%%%%%%%%%%
%%%%%%%%%%
%   Method
%%%%%%%%%%
\section{Data and Methodology}
\label{sec:method}

This section is organized as follows. Sections~\ref{sec:s1gal_sample} and \ref{sec:s2gal_sample} describe the galaxy and group samples used in this study, and Section~\ref{subsec:Spin} outlines the method employed to quantify the spin of galaxy clusters.

\subsection{Galaxy and Group Samples}

To address the potential impact of projection effects and to assess the robustness of our spin measurements, we make use of group catalogs constructed with two different methodologies, as summarized below.

Numerous group catalogs have been constructed from redshift samples obtained by the Sloan Digital Sky Survey (SDSS), using various algorithms such as the C4 algorithm \citep{2005AJ....130..968M}, friends-of-friends \citep[FoF,][]{2006ApJS..167....1B}, halo-based group finders \citep{2007ApJ...671..153Y}, and density-based methods \citep{,,,2023A&A...675A.161G}.  In particular, FoF-based catalogs are known to be susceptible to projection effects, which may bias measurements of galaxy cluster angular momentum. To minimize the influence of any single group finder and to perform an explicit consistency check, we therefore employ both an FoF-based group catalog as Sample-1 \citep{2017A&A...602A.100T} and a halo-based group catalog as Sample-2 \citep{2012ApJ...752...41Y}. The latter follows the halo-based method originally developed by \citet{2007ApJ...671..153Y}, which has been shown to effectively mitigate projection effects. As we demonstrate in the following sections, the spin signals measured from these two independent samples are consistent with each other, supporting the robustness of our results against the choice of group finder.

\subsubsection{Sample-1}
\label{sec:s1gal_sample}

For our Sample-1,
we use the spectroscopic galaxy catalog presented by \citet{2017A&A...602A.100T}, based on data from the Sloan Digital Sky Survey Data Release 12 \citep{2015ApJS..219...12A}.
To ensure data quality, the catalog was cleaned by applying a Petrosian $r$-band magnitude limit of $m_r = 17.77$~mag, effectively removing spurious or poorly measured entries. The resulting sample consists of 584,449 galaxies, each with group ID assignments, CMB-frame redshifts, and various diagnostic parameters.

The corresponding group catalog was constructed from the same galaxy sample using the Friends-of-Friends (FoF) algorithm \citep{1982ApJ...257..423H, 1982Natur.300..407Z, 1985ApJ...295..368B, 2009MNRAS.399..497D}. This algorithm links neighboring galaxies based on a defined projected linking length, grouping them into coherent systems. To account for redshift-dependent projection effects, the linking length was scaled with distance following the procedure described in \citet{2014A&A...566A...1T}. Group memberships were further refined according to the method detailed in \citet{2016A&A...588A..14T}, including a check for candidate merging systems, and group centre was calculated as the geometrical centre of all member galaxies without any luminosity or mass weighting as described in \citet{2016A&A...588A..14T}. The final catalog comprises 88,662 galaxy groups containing a total of 287,245 galaxies. Each group is described by its richness, Cartesian coordinates, and group ID.

\subsubsection{Sample-2}
\label{sec:s2gal_sample}

For our Sample-2, we select clusters from the catalog of \citet{2012ApJ...752...41Y}, which is constructed from spectroscopic galaxy data in the New York University Value-Added Galaxy Catalog (NYU-VAGC; \citealt{2005AJ....129.2562B}), itself based on SDSS DR7 \citep{2009ApJS..182..543A}. In the NYU-VAGC-based sample of \citet{2012ApJ...752...41Y}, galaxies in the Main Galaxy Sample with extinction-corrected apparent magnitude $r < 17.72$, redshifts in the range $0.01 \leq z \leq 0.20$, and redshift completeness $C_z > 0.7$ are selected, resulting in a spectroscopic sample of 639,359 galaxies covering 7,748 square degrees.

The corresponding group catalog was constructed from the same galaxy sample using the halo-based group finder of \citet{2007ApJ...671..153Y}. This group-finding algorithm is iterative and relies on an adaptive filter modeled after the general properties of dark matter halos. In contrast to the traditional FoF method, it is capable of identifying groups with only a single member, and the group center is determined as the luminosity-weighted geometric center of its member galaxies. The final catalog comprises 472,416 galaxy groups containing a total of 639,359 galaxies. Each group is characterized by its celestial coordinates, richness, and group ID.

%%%%%%%%%%%%% 
%Figure 1 
%%%%%%%%%%%%%
\begin{figure}[!ht]
\plotone{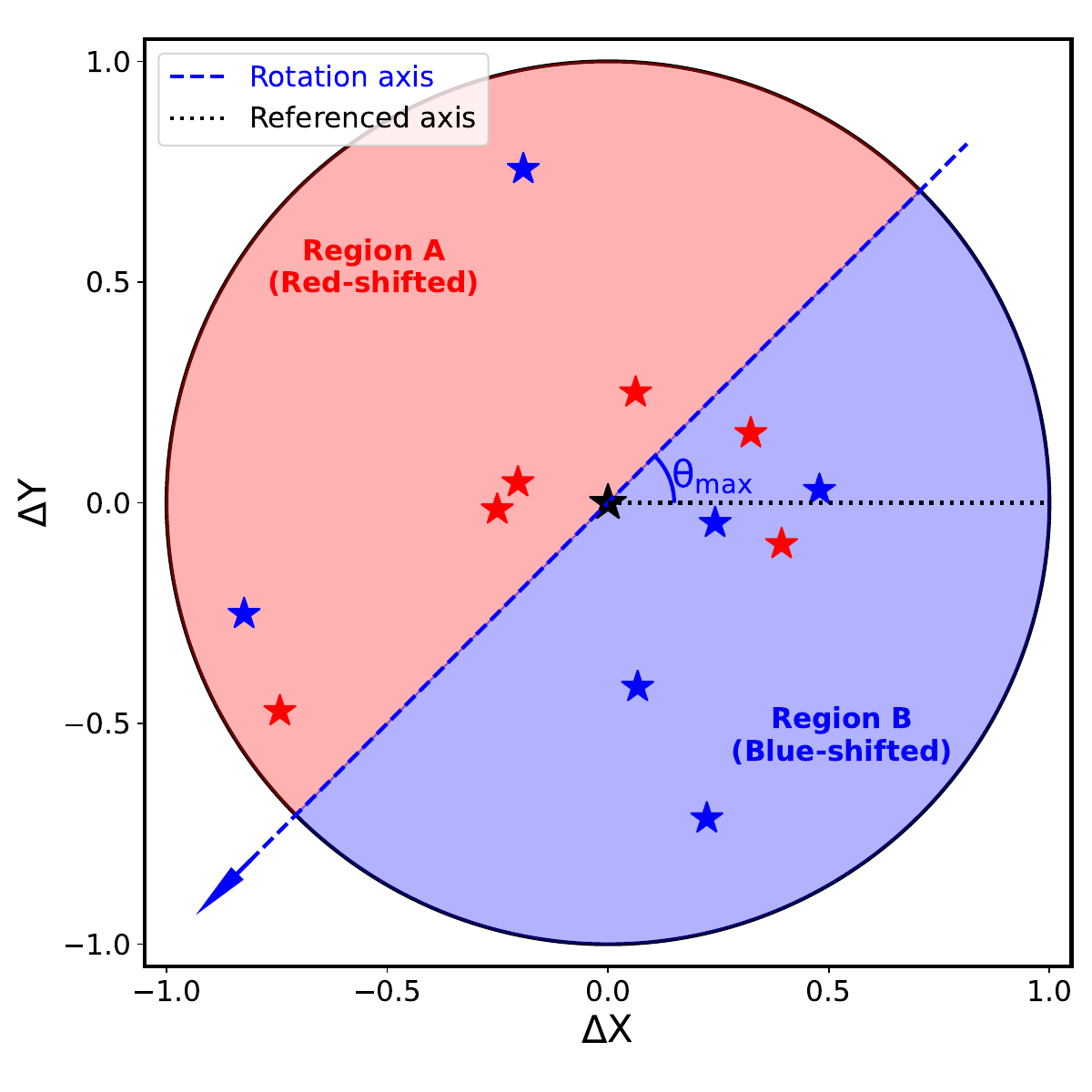}
\caption{A schematic illustration of the method for identifying the projected rotation axis of a galaxy cluster. Star symbols represent the the geometrical centre of all galaxies in the group(black) and the member galaxies (red and blue). The dashed line marks the projected rotation axis, which divides the projected plane into two regions: Region A (red) and Region B (blue). If the cluster is rotating in the direction indicated by the blue arrow, member galaxies in Region A are expected to predominantly exhibit red-shifted velocities, while those in Region B are expected to appear mostly blue-shifted.In this illustration, the projected rotation axis is rotated by an angle $\theta_{\rm max}$ from the reference axis (the x-axis, dotted line); see the main text for further details.}
\label{fig:cartoon}
\end{figure}

%%%%%%%%%%%%% 
%Figure 2 
%%%%%%%%%%%%%
\begin{figure*}[!ht]
\plotone{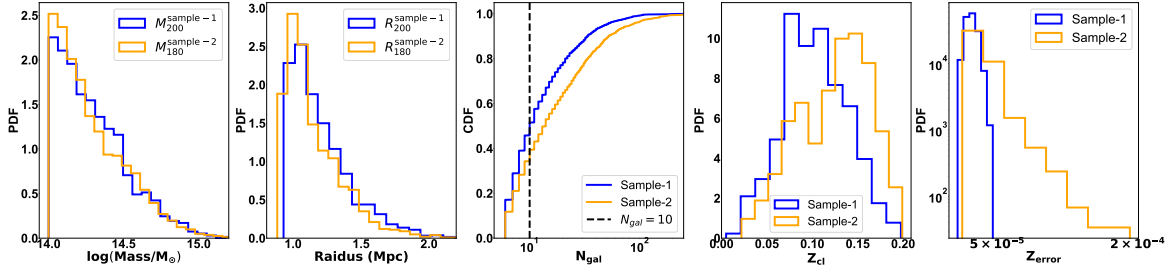}
    \caption{Distributions of the basic properties of the galaxy clusters used in this study for both Sample-1 (blue line) and Sample-2 (orange line). From left to right, the panels show: (1) cluster mass $\log(M_{\mathrm{cl}}/M_\odot)$; (2) cluster radius $R_{\mathrm{cl}}$ (in Mpc); (3) number of member galaxies $N_{\mathrm{gal}}$ (richness); (4) cluster redshift $Z_{\mathrm{cl}}$; and (5) spectroscopic redshift error $Z_{\mathrm{error}}$.}
    \label{fig:basic_properties}
\end{figure*}

\subsection{Identification of Cluster Spin}
\label{subsec:Spin}

To identify coherent rotation in galaxy clusters, we adopt a method inspired by techniques used to measure rotation in spiral galaxies via the Doppler shift of stellar populations. In our case, redshifts of member galaxies are analyzed to search for systematic differences indicative of bulk rotational motion. Notably, similar methodologies have been employed in studies of the spin of cosmic filaments \citep[e.g.,]{2021NatAs...5..839W, 2025ApJ...983..100W, 2025ApJ...982..197T}

A schematic illustration of this procedure is shown in the Figure~\ref{fig:cartoon}. Clusters are projected onto the X-Y plane, and similar results are obtained when projecting onto other planes. The initial trial rotation axis (i.e., the referenced axis) is taken to be the x-axis (the dotted line). This axis is then rotated from $0^\circ$ to $180^\circ$ in $1^\circ$ increments. For each trial orientation, the galaxies in a cluster are divided into two regions (labeled A and B) on either side of the axis, based on their projected positions. The mean redshifts of galaxies in regions A and B are then computed, denoted as $Z_A$ and $Z_B$, respectively. To ensure statistical significance, each region must contain at least three galaxies to be considered. Therefore, the minimum total number of galaxies required for a cluster is six.

The trial axis yielding the maximum redshift difference between the two regions is selected as the best-fit rotation axis of the cluster. The maximum redshift difference is then calculated as:
\begin{align}
\Delta Z_{\rm max} &= \max(|(Z_A - Z_0) - (Z_B-Z_0)|) \nonumber \\
                   &= \max(|Z_A - Z_B|)
\label{equ:dzab}
\end{align}
in which $Z_A$, $Z_B$ and $Z_0$ are the mean redshifts of galaxies in region A, region B, and all member galaxies of the cluster, respectively.
It is worth noting that, according to the above formula, whether $Z_0$ is taken as the mean redshift of all member galaxies of the cluster or the redshift of the central galaxy does not affect the value of $\Delta Z_{\rm max}$.
The $\Delta Z_{\rm max}$ serves as a proxy for the cluster's spin amplitude. The value of $\theta_{\rm max}$ is then used to determine the orientation of the rotation axis of the cluster in the projected plane. 
It is worth noting that, in the process of determining $\theta_{\rm max}$ as shown in Figure~\ref{fig:cartoon}, we sometimes encounter cases where the same value of $\Delta Z_{\rm max}$ corresponds to multiple values of $\theta_{\rm max}$. This phenomenon is caused by the sparse spatial distribution and the limited number of member galaxies, as discussed in the Appendix~\ref{app:sec1}(corresponding to Sample-1). In addition, the root-mean-square (RMS) deviation of the redshifts of member galaxies, denoted as $Z_{\rm rms}$, is also calculated to quantify the redshift dispersion within each cluster. We further adopt the ratio $Z_{\rm rms}/\Delta Z_{\max}$ as a proxy for the cluster dynamical ``temperature'', similar to the filament dynamical temperature defined in \cite{2021NatAs...5..839W}.

To assess the statistical significance of the detected spin signal, we perform a Monte Carlo randomization test, which was alsoalso used in \cite{2021NatAs...5..839W,2025ApJ...983..100W,2025ApJ...982..197T}.
\begin{itemize}
    \item Shuffle the redshifts of the galaxies in each cluster while keeping their positions fixed. 
    \item Calculate $\Delta Z_{\rm max}^{\rm rand}$ by, for each Monte Carlo realization, finding the rotation axis that maximizes $\Delta Z_{\rm max}$ and recording that maximum value. the same selection procedure is applied to the randomized samples as to the real cluster, ensuring fair comparison.
    \item Repeat this procedure 10,000 times for each cluster.
    \item Compute the significance of each observed cluster $\Delta Z_{\rm max}$ using:
    \begin{equation}
        s = \frac{\Delta Z_{\rm max} - \langle \Delta Z_{\rm max}^{\rm rand} \rangle}{\sigma(\Delta Z_{\rm max}^{\rm rand})},
    \end{equation}
    where $\langle \Delta Z_{\rm max}^{\rm rand} \rangle$ and $\sigma(\Delta Z_{\rm max}^{\rm rand})$ are the mean and standard deviation over the 10,000 randomized realizations.
\end{itemize}

Following the conventional definition of galaxy clusters, we select systems with total masses $M > 10^{14}\, M_\odot$ from the two group catalogs.
The resulting sample comprises 2,170 and 1329 galaxy clusters respectively, which are used for the following analysis of cluster spin and compared against randomized control samples. 
The basic statistical properties of the cluster samples, including virial mass $M$, virial radius $R$, richness $N_{\mathrm{gal}}$, and redshift $z$, are summarized in Figure~\ref{fig:basic_properties}. The similar distributions of the basic properties of the two samples ensure the feasibility of a comparative analysis, particularly with respect to mass and radius. The slight differences in $N_{\mathrm{gal}}$ and redshift $z$ mainly arise from the different galaxy input catalogs and the distinct group-finding algorithms adopted in the two cases.
In the third panel, we can see that for sample-1 is roughly divided into two equal subsamples at $N_{\mathrm{gal}} = 10$. This division will be taken into account in the subsequent analysis. The corresponding richness distributions for both samples are also displayed in the left panel of Figure~\ref{fig:appendix_2} for reference.
In addition, the last panel shows that the peak of the spectroscopic redshift error occurs at roughly $2.5 \times 10^{-5}$ for both samples, while Sample-2 exhibits a somewhat broader range, with a maximum reaching only $\sim 2 \times 10^{-4}$.
Part of the difference arises from the known quantization of inverse variance in the SDSS DR7 pipeline, which artificially increases the reported flux uncertainties in low-S/N regions; this issue is largely mitigated in the DR12 reduction, leading to systematically smaller and smoother error estimates.
which means that, based on the method described above, any $\Delta Z_{\rm max}$ less than or equal to $z_{\rm error}$ is not meaningful.

%%%%%%%%%%%%%%%%%%%%%%%%%%%%%%%%%%%%%%%%%%%%%%%%%%%%%
%%%%%%%%%%%%% 
%   Result
%%%%%%%%%%%%% 
\section{Result}
\label{sec:result}

\begin{figure*}[!htp]
\plotone{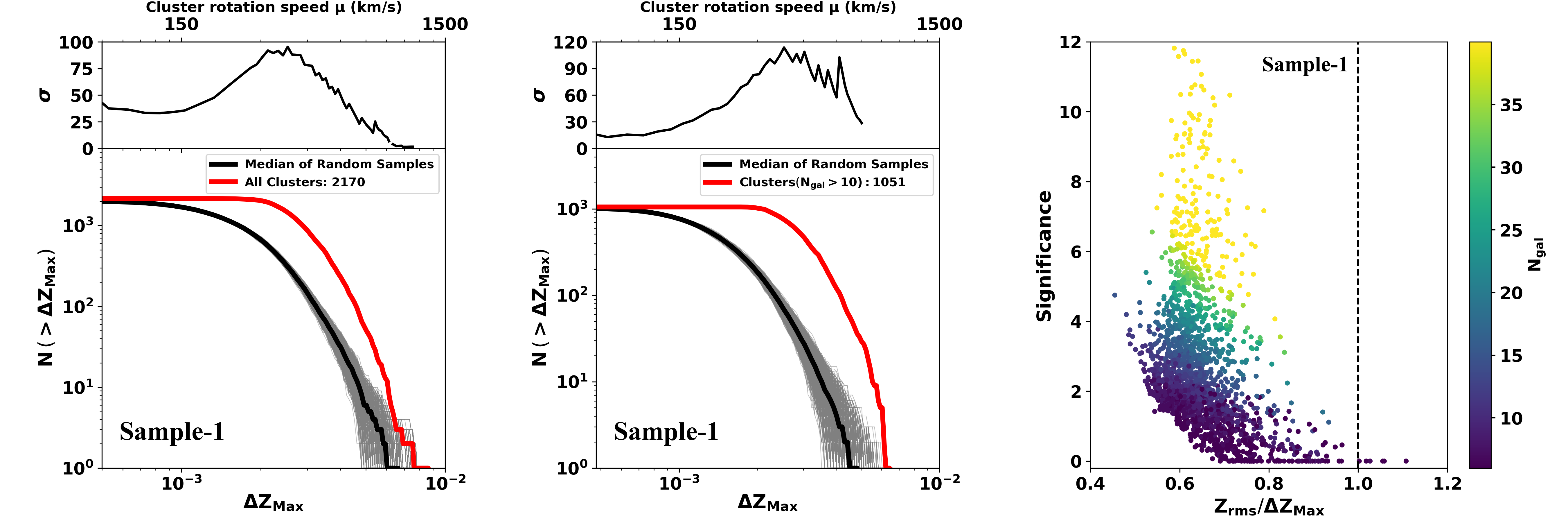}
\plotone{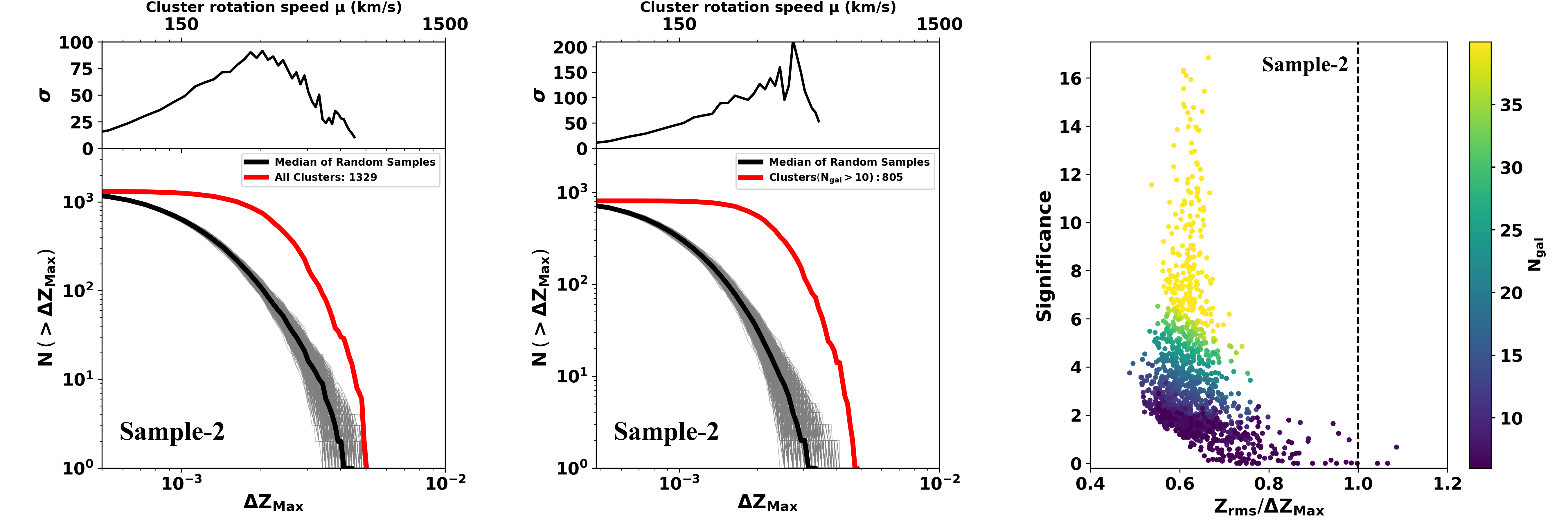}
  \caption{The global detection of cluster spin for Sample-1 (top row) and Sample-2 (bottom row).
  \textbf{Left:} The cumulative distribution of the redshift difference, $\Delta Z_{\rm max}$, between galaxies in region A and region B. The solid red lines in the bottom panel represent the distribution of the measured cluster spin signal, while the 10,000 gray lines correspond to randomized trials. The black solid lines indicate the median values of the random samples. The top panels display the sample-to-sample distance between the real clusters in the observation and the random samples, expressed in units of the standard deviation of the random trials. The upper x-axes show the rotation speed of the filaments, calculated as $\mu=\frac{1}{2}c\times\dzmax$.
  \textbf{Middle:} Similar to the left panel, but show clusters with member galaxies more than 10 ($N_{gal}>10$).
  \textbf{Right:} The significance of the cluster spin signal as a function of the ratio $Z_{\mathrm{rms}}/\Delta Z_{\rm max}$. Each point represents a cluster, colored by the number of member galaxies in regions A and B ($N_{\mathrm{gal}}^{A,B}$). The vertical dashed line marks $Z_{\mathrm{rms}}/\Delta Z_{\rm max} = 1$.}
    \label{fig:fig3}
\end{figure*}

To investigate the presence of coherent rotational motion in galaxy clusters, we analyze the maximum redshift difference, $\Delta Z_{\rm max}$, between galaxies situated in two diametrically opposed regions, A and B, as illustrated in Figure~\ref{fig:cartoon}. The main results of this analysis are presented in the following three subsections, focusing on the global detection of cluster spin signals (Figure~\ref{fig:fig3}), the radial profile of cluster spin (Figure~\ref{fig:fig4}), and the alignment of cluster spin with central galaxy and large-scale structures (Figure~\ref{fig:fig5}).

%%%%%%%%%%%%% 
%Figure 3
%%%%%%%%%%%%%

\subsection{Global Detection of Cluster Spin Signals}
The measured coherent spin signals of galaxy clusters for the two samples are presented in Figure~\ref{fig:fig3}, with the top row corresponding to Sample-1 and the bottom row to Sample-2. 
Qualitatively consistent results are obtained for the two cluster samples defined by different group-finding algorithms. The observed cumulative distributions show a consistent and significant excess at higher values of $\Delta Z_{\rm max}$ compared to the random samples, further indicating that the choice of algorithm does not affect the measurement of the spin signal and validating the robustness and reliability of our method. Some quantitative differences between the two samples are present.
For example, the signal reaches its maximum confidence level at $\sim 380\ \mathrm{km\,s^{-1}}$ for Sample-1 and at $\sim 300\ \mathrm{km\,s^{-1}}$ for Sample-2, which we attribute to differences in the underlying galaxy samples and in the cluster-finding algorithms. In the following, we focus on the results from Sample-1 for illustrative purposes, noting that Sample-2 yields consistent trends with slightly different amplitudes.

The left panel of Figure~\ref{fig:fig3} presents the cumulative distribution of $\Delta Z_{\rm max}$ for the observed clusters, compared against a suite of 10,000 randomized control samples.
In the bottom sub-panel, the red solid line shows the cumulative distribution of the measured signal, while the gray lines represent the distributions from the randomized trials, where galaxy redshift have been shuffled but galaxy position are kept to erase any intrinsic rotation. The black solid line marks the median of the random samples, serving as a baseline for comparison. To assess the statistical significance of the observed signal, the top sub-panel displays the sample-to-sample deviation between the observed distribution and each random realization, expressed in units of the standard deviation derived from the random trials. This metric quantifies how strongly the observed data deviates from expectations under the null hypothesis of no rotation. For interpretability, the upper x-axis translates the redshift difference into an effective rotational velocity using  
\begin{equation}
\mu = \frac{1}{2}c \times \frac{\Delta Z_{\rm max}}{1+Z_{\rm cl}},
\label{eq:mu_def}
\end{equation}
where $c$ is the speed of light and $\Delta Z_{\rm max}$ 
is the measured maximum redshift difference between the two regions, $Z_{\mathrm{cl}}$ is the cluster redshift, calculated as an average over all cluster members. The observed cumulative distribution exhibits a consistent and significant excess at higher values of $\Delta Z_{\rm max}$ compared to the random samples. 
As shown in the top sub-panel, the cumulative deviation reaches a level corresponding to an apparent $\sim100\sigma$ when integrated over all clusters (around $\sim380\,\mathrm{km\,s^{-1}}$). This value reflects the aggregated, sample-to-sample confidence rather than the significance of any individual cluster. The deviation remains substantial across a wide range of $\Delta Z_{\rm max}$, indicating a statistically robust collective trend of coherent rotational motion among the observed clusters.
For completeness, the probability density distribution (PDF) of $\Delta Z_{\rm max}$ is displayed in the middle panel of Figure~\ref{fig:appendix_2}, illustrating the detailed shape of the distribution.

The middle panel shows the same analysis, but restricted to clusters with more than 10 member galaxies ($N_{\mathrm{gal}} > 10$). The results are qualitatively similar, with the observed clusters again displaying a significant excess in $\Delta Z_{\rm max}$ relative to the randomized controls. 
Notably, the cumulative sample-to-sample deviation shown in the upper sub-panel is even more pronounced than in the full sample, reaching an apparent level of $\sim120\sigma$ at $\sim379\,\mathrm{km\,s^{-1}}$. This value represents an aggregated confidence across richer clusters rather than the significance of individual systems, indicating a stronger collective rotational tendency in these high-richness clusters.
The right panel displays the significance of the cluster spin signal as a function of the ratio $Z_{\mathrm{rms}}/\Delta Z_{\rm max}$ for individual clusters, where the x-axis represents the dynamical temperature as introduced in the study of filament spin by \cite{2021NatAs...5..839W}. Each point corresponds to a cluster and is colored according to the number of member galaxies in regions A and B ($N_{\mathrm{gal}}^{A,B}$). The vertical dashed line marks $Z_{\mathrm{rms}}/\Delta Z_{\rm max} = 1$. Two key trends are evident: (1) clusters with lower dynamical temperatures (i.e., smaller $Z_{\mathrm{rms}}/\Delta Z_{\rm max}$ ratios) tend to exhibit higher significance, indicating that the spin signal is more prominent in dynamically colder systems; and (2) at a fixed dynamical temperature, clusters with a larger number of member galaxies (indicated by the yellow color) show higher significance, suggesting that richer clusters provide a more robust detection of coherent rotation. 

We also performed Kolmogorov--Smirnov (KS) tests to quantify the differences between the observed sample and the randomized realizations. For the direct two-sample comparison, we obtained a KS statistic of $D = 0.6986$ with a $p$-value $p \approx 0.0\mathrm{e}{+}00$ (i.e., $p \ll 10^{-300}$). For the comparison of the corresponding cumulative distributions (the red and black curves in Figure~\ref{fig:fig3}), we found $D = 0.305$ with $p = 1.87 \times 10^{-41}$. These extremely small $p$-values indicate that the observed and randomized samples are statistically inconsistent, consistent with the clear visual separation between the red and black curves in Figure~\ref{fig:fig3}.

\subsection{Radial Profile of Cluster Spin}

Building on the statistical evidence for cluster spin presented in Figure~\ref{fig:fig3}, we now turn to the spatial distribution of galaxies within clusters to further elucidate the observed rotational signal.

\begin{figure*}[!ht]
\centering
\plotthree{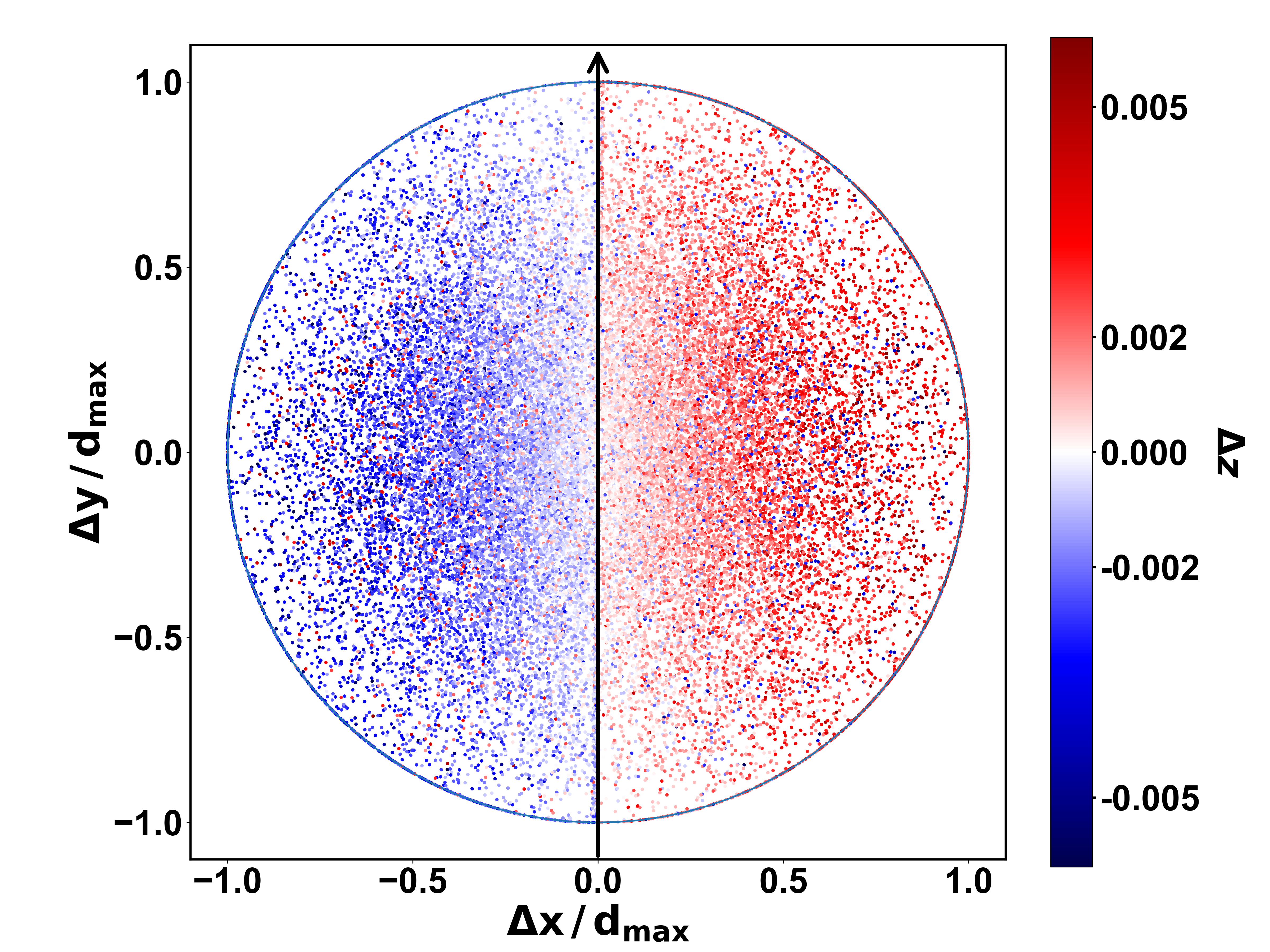}{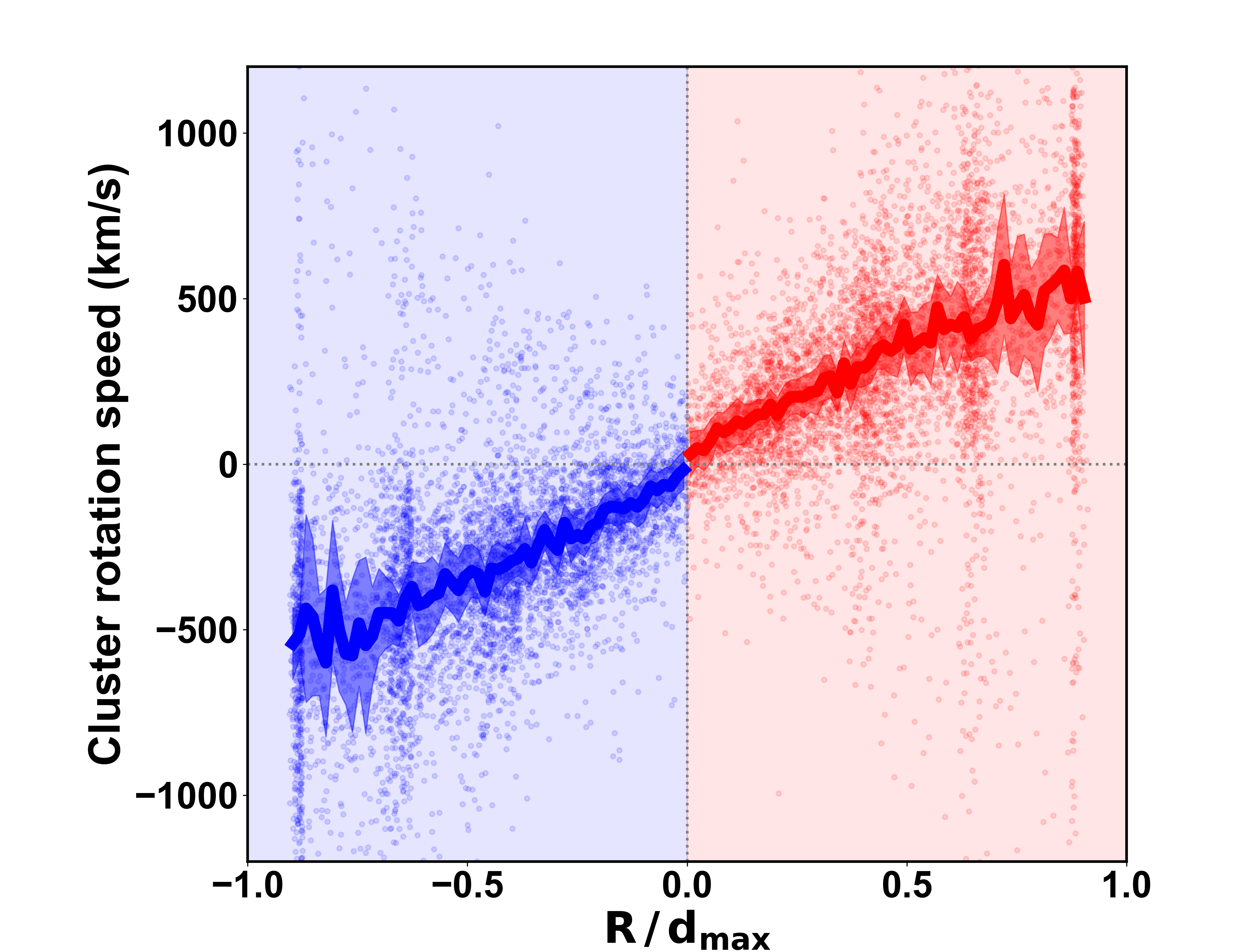}{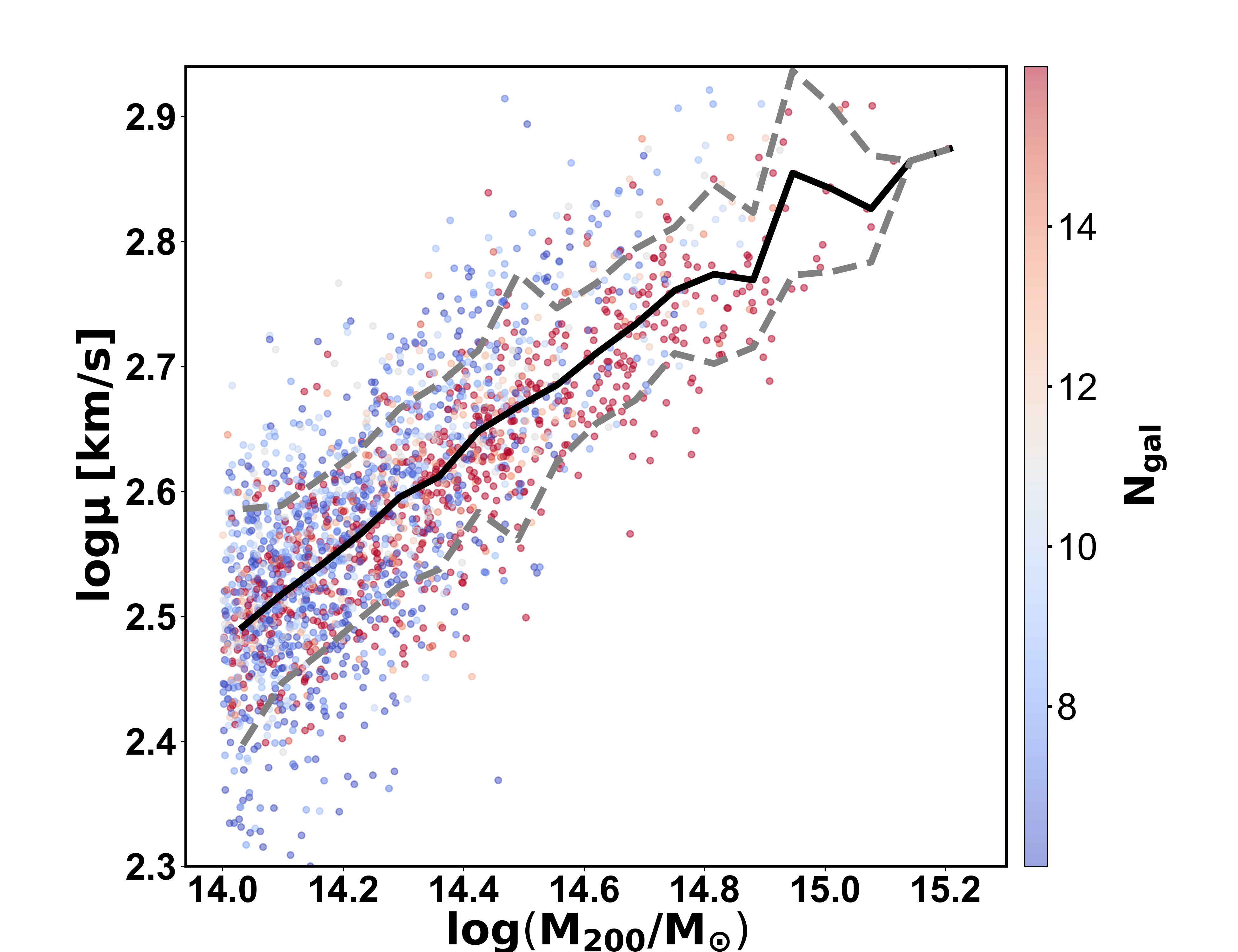}
\caption{\textbf{Left:} Projected distribution of galaxies within clusters, where the $x$- and $y$-axes represent normalized coordinates ($\Delta x/d_{\mathrm{max}}$, $\Delta y/d_{\mathrm{max}}$) centered on the cluster center, with $d_{\mathrm{max}}$ defined as the distance to the most distant member galaxy in each cluster. Each point is color-coded by its redshift deviation $\Delta z$ from the cluster mean; redder (bluer) points indicate galaxies moving away (approaching). The black vertical line marks the rotation axis, and the arrow shows the inferred spin direction. All clusters are rotated to align their spin axes vertically. 
\textbf{Middle:} Cluster rotation speed as a function of normalized distance to the center. Each point shows the measured rotation speed of a member galaxy, with blue and red points for galaxies on opposite sides of the rotation axis. Distances in the receding (approaching) region are shown in red (blue) and assigned positive (negative) values. 
\textbf{Right:} The relation between circular velocity $\log \mu$ (km\,s$^{-1}$) and cluster mass, $\log(M_{\rm 200}/M_\odot)$. Each data point represents a single cluster, color-coded by its richness, $N_{gal}$, as indicated by the colorbar. The black solid line shows the median relation, while the grey dashed lines represent the $1\sigma$ scatter in each mass bin.}
\label{fig:fig4}
\end{figure*}

Following the schematic illustration of the spin for a single cluster shown in Figure~\ref{fig:cartoon}, we stack the signals to intuitively inspect the spin visualization. The left panel of Figure~\ref{fig:fig4} displays the projected distribution of galaxies within galaxy clusters. The $x$- and $y$-axes correspond to the normalized coordinates $\rm\Delta x/d_{max}$ and $\rm\Delta y/d_{max}$, where $\rm d_{max}$ denotes the largest distance of a member galaxy from the cluster center. Each point represents a galaxy, color-coded by its redshift deviation $\Delta z$ relative to the cluster mean redshift, as indicated by the colorbar. Redder points correspond to galaxies with positive $\Delta z$ (redshifted, moving away), while bluer points correspond to negative $\Delta z$ (blueshifted, approaching).

The black vertical line marks the identified axis of rotation for each cluster, and the arrow indicates the direction of the inferred spin, with the left side being blueshifted and the right side being redshifted. For visualization purposes, all clusters are rotated such that their spin axes are aligned vertically. This alignment allows for a direct comparison of the spatial and kinematic structures across the cluster sample. It is worth noting that the redshifted (red, on the right) and blueshifted (blue, on the left) regions are not entirely pure; blue points are embedded within the red regions and vice versa, reflecting the inherent complexity and scatter in the observational data.

The spatial segregation of redshifted and blueshifted galaxies on opposite sides of the rotation axis, as seen in the left panel of Figure~\ref{fig:fig4}, provides direct visual evidence for coherent rotational motion within clusters. This pattern is fully consistent with the statistical signal detected in Figure~\ref{fig:fig3}, reinforcing the conclusion that the observed $\Delta Z_{\rm max}$ distribution arises from genuine physical rotation rather than random fluctuations. 

To further quantify the rotational motion suggested by the spatial segregation in the left panel, the middle panel of Figure~\ref{fig:fig4} presents the cluster rotation speed at given radius as a function of normalized distance from the cluster center. Here, Each point represents the measured galaxy rotation speed at a given distance from the cluster center, normalized by $d_{\rm max}$. Blue and red points correspond to galaxies on opposite sides of the rotation axis. Notably, the rotation speed increases with radius. The scatter becomes larger at greater radius, primarily because the number of galaxies decreases with increasing distance from the cluster center.

In the right panel of Figure~\ref{fig:fig4}, we examine the rotation speed $\mu$ of galaxy clusters, defined in Equ.~\ref{eq:mu_def}, which corresponds to the upper $x$-axis in Figure~\ref{fig:fig3}, as a function of the cluster virial mass. Each data point represents an individual cluster, with the color indicating the number galaxies of the clusters, ranging from smaller (blue) to larger (red) clusters. The solid line shows the mean distribution, while the dashed lines indicate the $1\sigma$ statistical uncertainty range. It can be seen that as the cluster mass increases, the rotation speed rises from $\sim 360\,\mathrm{km\,s^{-1}}$ at $M_{\rm vir} \sim 10^{14}\,M_\odot$ to $\sim 693)\,\mathrm{km\,s^{-1}}$ at $M_{\rm vir} \sim 10^{15}\,M_\odot$. This positive correlation suggests that more massive clusters tend to exhibit stronger rotational motion. 
The distribution of the red data points (corresponding to larger values of $N_{\mathrm{gal}}$) generally follows the same trend as that of the blue data points (with smaller $N_{\mathrm{gal}}$), but is more tightly clustered around the black solid line. This indicates that, although the high-$N_{\mathrm{gal}}$ systems are fewer in number, they dominate the relation between rotation speed $\mu$ and cluster mass.

%%%%%%%%%%%%% 
%Figure 4
%%%%%%%%%%%%%

\subsection{Alignment of Cluster Spin}

Having established the global and radial profile of coherent spin in galaxy clusters, we now turn to the question of how cluster spin we measured is oriented relative to other key directions, both central galaxy within the cluster and in the surrounding large-scale structure. Specifically, we examine the alignment between the cluster spin axis and the spin of the central galaxy, as well as the angle between the cluster spin and the direction of the nearest cosmic filament. This serves as an initial exploration of how angular momentum is transferred from large-scale structures (Mpc scale) to the cluster scale and ultimately to the central galaxy (Kpc scale).

Early work on the alignment of central cluster galaxies and their host clusters has been extensively studided. The observational support for central galaxy alignment is solid and uncontested, Most studies determine the presence and dependence of alignment by comparing the major axis of the central galaxy's shape (elliptical fit) with that of the host cluster (calculated from the distribution of member galaxies using inertial momentum or by fitting the X-ray gas distribution)(e.g., \citealt{1968PASP...80..252S,1982A&A...107..338B,2010MNRAS.405.2023N,2016MNRAS.463..222H,2017NatAs...1E.157W,2019ApJ...874...84W,2022MNRAS.516.3159Y}). By specifying the rotation axes of the clusters in our method, we can further supplement and validate the alignment between the central galaxy and the host cluster.

\begin{figure*}[!ht]
\centering
\plotone{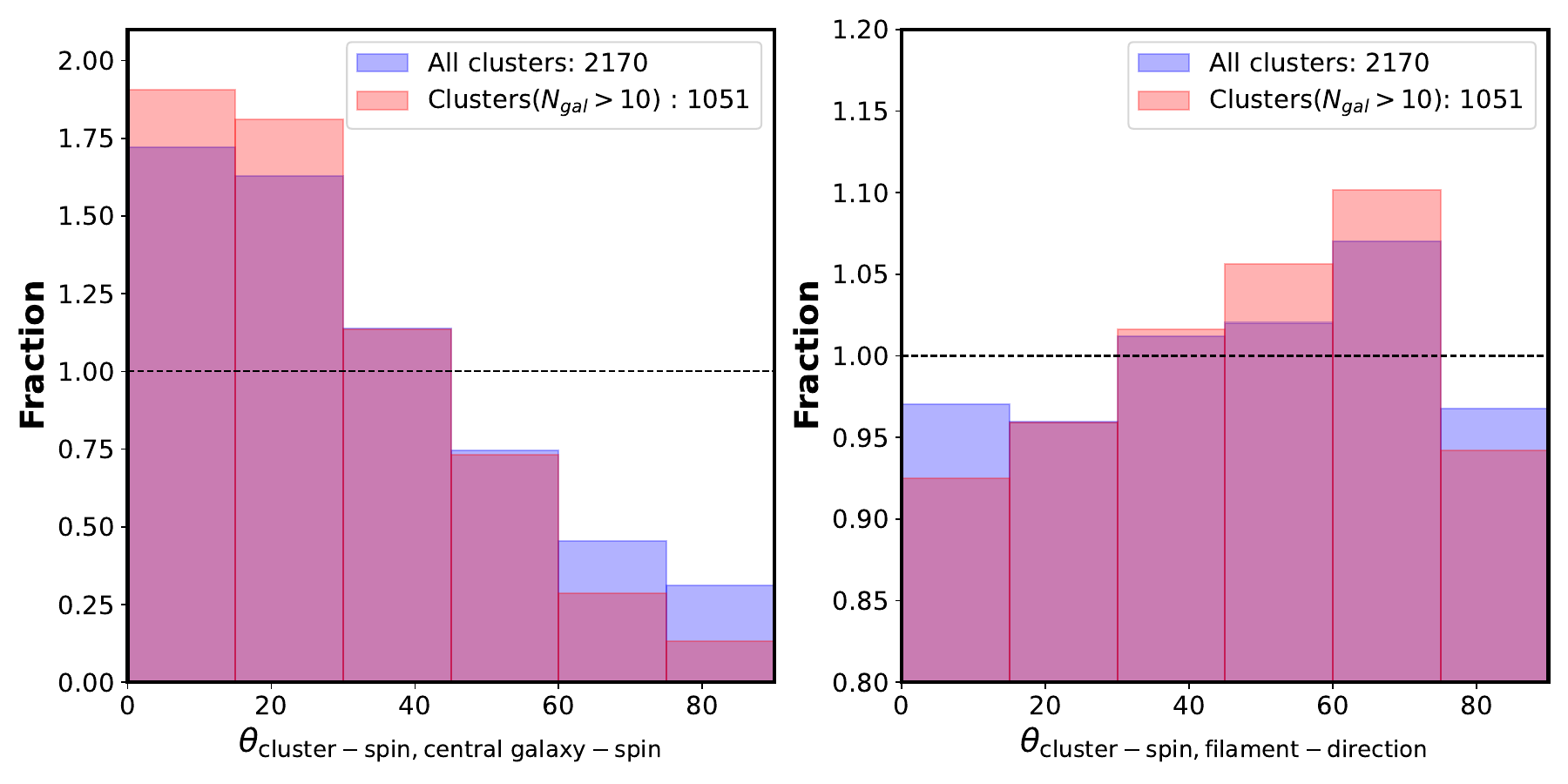}
\caption{\textbf{Left:} Distribution of the angle between the cluster spin axis and the spin of the central galaxy. The blue and red histograms represent all clusters and clusters with more than 10 member galaxies ($N_{\mathrm{gal}} > 10$), respectively. 
\textbf{Right:} Distribution of the angle between the cluster spin axis and the orientation of the nearest cosmic filament. The blue and red histograms represent the same samples as in the left panel.}
\label{fig:fig5}
\end{figure*}

Figure~\ref{fig:fig5} presents the results of this alignment analysis. In the left panel, we show the distribution of the angle between the cluster spin axis and the spin of the central galaxy. 
Here, the spin of the central galaxy is calculated following the methods of \citet{2007ApJ...671.1248L, 2012ApJ...744...82V, 2021MNRAS.504.4626K, 2015ApJ...798...17Z, 2025ApJ...987L..30W}, using the observed position angle, inclination, and sky coordinates. 
The blue line represents the full cluster sample, while the red line corresponds to clusters with more than 10 member galaxies ($N_{\mathrm{gal}} > 10$). 
Both samples exhibit a clear tendency for parallel alignment between the cluster spin and the central galaxy spin, with the alignment signal being notably stronger in the richer ($N_{\mathrm{gal}} > 10$) subsample. Recently, \citet{2025ApJ...992L..17W} investigated in detail the alignment between the spin axes of galaxy clusters and their central galaxies, finding that the alignment strength depends on both the cluster mass and the mass of the central galaxy. Our results are consistent with their findings.

The right panel displays the distribution of the angle between the cluster spin axis and the orientation of the nearest filament. To determine the filament orientation, we identify the closest filament to each cluster using the filament catalog used in \cite{2021NatAs...5..839W} with Bisous model \citep{2014MNRAS.438.3465T}, and adopt the local filament orientation as the reference axis. The results show a preference for perpendicular signal between the cluster spin and the filament orientation, with this signal being more pronounced in clusters with $N_{\mathrm{gal}} > 10$. This finding is consistent with recent observational \citep{2025ApJ...983L...3R} and simulation \citep{2025JCAP...10..095W} results.

%%%%%%%%%%%%% 
%Figure 5
%%%%%%%%%%%%%

%%%%%%%%%%%%%%%%%%%%%%%%%%%%%%%%%%%%%%%%%%%%%%%%%%%%%
%%%%%%%%%%%%% 
%   con & dis
%%%%%%%%%%%%% 
\section{Summary}\label{sec:sum}
%summarize this work
In this study, we investigated the spin properties of galaxy clusters using FoF-based group catalog \citep{2017A&A...602A.100T} and the halo-based group catalog \citep{2012ApJ...752...41Y}. Our cluster samples consists of 2,170 and 1329 systems with $M > 10^{14}\, M_\odot$, respectively.
We quantified the spin of each cluster by searching for the orientation in the projected plane that maximizes the redshift difference between two hemispheres of member galaxies, and assessed the statistical significance of the detected spin signals using Monte Carlo randomizations.

Our results provide strong statistical evidence for the presence of coherent rotational motion in galaxy clusters. The observed distribution of the maximum redshift difference, $\Delta Z_{\rm max}$, between member galaxies on opposite sides of the projected rotation axis, shows a significant excess compared to randomized control samples, with a confidence level reaching its peak at $\sim 380 \kms$ for Sample-1 and at $\sim 300 \kms$ for Sample-2. This signal is even more pronounced in richer clusters with more than 10 member galaxies. Stacked visualizations further confirm that the projected rotation axes effectively separate redshifted and blueshifted galaxy populations within clusters.

We also examined the radial profile of the spin signal, finding that the rotation speed increases with cluster mass, from approximately $330~\mathrm{km\,s}^{-1}$ at $M_{\rm vir} \sim 10^{14} M_\odot$ to about $800~\mathrm{km\,s}^{-1}$ at $M_{\rm vir} \sim 10^{15} M_\odot$. Additionally, we explored the alignment of cluster spin with both the spin of the central galaxy and the orientation of the nearest cosmic filament. Our analysis reveals a tendency for the cluster spin axis to be parallel to the central galaxy spin, and preferentially perpendicular to the direction of the nearest filament, especially in richer clusters.

These findings demonstrate that galaxy clusters exhibit statistically significant and physically coherent spin signals, which are linked to both their internal properties and their large-scale environment. The origin and evolution of cluster spin, and its connection to cluster dynamics and cosmic structure, will be further explored in future work.

\section{Discussion}\label{sec:dis}

As illustrated schematically in Figure~\ref{fig:cartoon}, our method offers a distinct advantage in that it not only quantifies the amplitude of the spin signal, but also provides a well-defined orientation for the projected angular momentum axis of each cluster. By maximizing the redshift difference between two regions divided by a trial axis, we are able to spatially separate redshifted and blueshifted galaxy populations within the cluster, thus assigning a direction to the spin axis in the plane of the sky. This directional information is particularly valuable, as it enables direct comparison with other independent tracers of cluster or galaxy angular momentum. Notably, recent observational work using MaNGA \citep{2017AJ....154...86W} data has, for the first time, revealed an anti-parallel alignment \citep{2025ApJ...987L..30W} between low-mass spiral galaxy spin directions and the spin of their host filament. Our method, which provides a measurement of the projected spin axis and its direction for each cluster, opens up new avenues for testing alignment signals at the cluster scale and for investigating the interplay between cluster dynamics and the surrounding large-scale structure.

In particular, future studies can combine our galaxy-based spin measurements with independent probes such as the kinematic Sunyaev-Zel'dovich (kSZ) effect and X-ray observations. Both the kSZ effect and X-ray data offer insights into the velocity structure of the intracluster medium, providing an independent view of cluster dynamics. By comparing the spin inferred from galaxy dynamics with that derived from the kSZ effect or X-ray gas motions, we can perform valuable cross-checks to validate and refine our spin measurement technique. Such multi-wavelength approaches will not only enhance the robustness of our results but also deepen our understanding of the dynamical state and angular momentum acquisition of galaxy clusters.

The origin of cluster angular momentum is intimately connected to the large-scale structure of the Universe \citep{2002ApJ...581..799V}. It is widely recognized that the majority of the mass accreted by galaxy clusters is funneled through the cosmic filaments \citep{2013MNRAS.428.2489L, 2014MNRAS.443.1274L,2015ApJ...813....6K} that connect to them, making these filaments the dominant channels for the buildup of cluster angular momentum. In the framework of hierarchical structure formation, galaxy clusters correspond to the most massive dark matter halos, whose angular momentum is thought to arise primarily from the anisotropic infall of matter along these filaments. Numerical simulations have consistently shown that the angular momentum vectors of massive halos tend to be oriented perpendicular \citep{2017MNRAS.468L.123W,2018MNRAS.473.1562W,2018MNRAS.481..414G,2019MNRAS.487.1607G,2021MNRAS.503.2280G} to the direction of the nearest filament. When using member galaxies as tracers to estimate the angular momentum of the underlying dark matter halo  \citep{ 2025JCAP...10..095W}, the resulting distribution is expected to reflect that of the halo itself.

Our results, together with recent observational findings \citep{2025ApJ...983L...3R}, indicate that the spin axes of clusters are preferentially perpendicular to the orientation of the nearest filament. 
As shown in the right panel of Figure~\ref{fig:fig5} and also reported by \citet{2025ApJ...983L...3R} and \cite{2025JCAP...10..095W}, this perpendicular signal between cluster spin and nearby filament is present but relatively weak.
This is consistent with a scenario in which the infall of matter along filaments imparts a net rotational motion to the cluster, establishing a preferred spin-filament alignment. In addition to filamentary accretion \citep{2018MNRAS.473.1562W}, clusters may also acquire angular momentum through the accretion of more diffuse material or isolated galaxies from the surrounding environment \citep{2014MNRAS.445L..46W}. The combined effects of these processes shape the observed spin properties of galaxy clusters and underscore the complex interplay between cluster growth and the cosmic web.

Understanding how angular momentum is transferred across cosmic scales is crucial not only for interpreting the spin properties of large-scale filament \cite{2021NatAs...5..839W, 2025ApJ...982..197T, 2025ApJ...983..100W}, galaxy clusters and their member galaxies, but also for the physical understanding of tidal torque theory (TTT) \citep{2002MNRAS.332..325P, 2002MNRAS.332..339P, 2021MNRAS.502.5528L, 2024PASP..136c7001L, 2025arXiv250501298L}. Our finding that the spin axis of the central galaxy tends to be parallel to the overall cluster spin suggests a coherent connection between the angular momentum acquired from the large-scale structure and that retained by the most massive galaxy at the cluster center. 
As shown in the left panel of Figure~\ref{fig:fig5}, there is a very strong parallel alignment between the cluster spin and the spin of the central galaxy, indicating that the central galaxy can efficiently inherit the angular momentum of the host cluster.
This parallel alignment suggests that angular momentum may be rapidly transferred from the cosmic web—primarily through filamentary accretion—first to the cluster as a whole, and subsequently to its central galaxy. Such a scenario is consistent with hierarchical structure formation, where the direction of infalling material along filaments not only sets the spin of the cluster but also influences the angular momentum of the central galaxy, either through direct accretion or through the cumulative effects of mergers and interactions within the cluster core. These results highlight the multi-scale nature of angular momentum acquisition, linking the orientation of cluster spin to both the surrounding large-scale environment and the internal structure of the cluster. The stronger alignment signals observed in richer clusters further support the idea that sustained and coherent accretion along filaments plays a dominant role in shaping the spin properties of both clusters and their central galaxies.

The detection of coherent rotational motion in galaxy clusters not only deepens our understanding of their formation and evolution, but also suggests a potential new avenue for cluster mass measurements. In principle, the spin signal traced by member galaxies could provide an independent estimate of cluster mass, complementing traditional methods such as those based on velocity dispersion \citep{2014A&A...566A...1T,2020ApJS..247...12S} or strong/weak gravitational lensing\citep{2015MNRAS.454.4085B,2022ApJ...928...87F}. While the practical implementation and calibration of such spin-based mass estimators require further investigation, this approach may offer a promising and complementary tool for future studies of cluster dynamics and mass determination. A detailed exploration of cluster mass measurements based on spin signals will be presented in our forthcoming work.

\begin{acknowledgments}
We thank Prof.\ Xi Kang (Zhejiang University) and Prof.\ Tao Wang (Nanjing University) for insightful discussions and comments.
PW acknowledge the financial support from the NSFC (No.12473009), and also sponsored by Shanghai Rising-Star Program (No.24QA2711100). This work is supported by the China Manned Space Program with grant no. CMS-CSST-2025-A03. 
Y.R. acknowledges supports from the CAS Pioneer Hundred Talents Program (Category B), the NSFC grants 12522302 and 12273037, and the USTC Research Funds of the Double First-Class Initiative. This work is also supported by the China Manned Space Program with grant no. CMS-CSST-2025-A06 and CMS-CSST-2025-A08.
M.B. acknowledges support by the National Natural Science Foundation of China, NSFC grant No. 12303009.
\end{acknowledgments}

\appendix
\setcounter{figure}{0}
\renewcommand{\thefigure}{A\arabic{figure}} % A1, A2, ...

\section{Ambiguity in the Determination of the Projected Rotation Axis}
\label{app:sec1}

In the process of determining the projected rotation axis, we occasionally find that a single value of $\Delta Z_{\rm max}$ can correspond to multiple values of $\theta_{\rm max}$. This phenomenon arises primarily due to the sparse spatial distribution and limited number of member galaxies within some clusters. When the number of member galaxies is small or their distribution is highly asymmetric, the calculation of $\Delta Z_{\rm max}$ as a function of the trial rotation axis angle $\theta$ may yield several local maxima of similar amplitude. As a result, multiple candidate angles $\theta_{\rm max}$ can be identified for the same $\Delta Z_{\rm max}$. 

The left panel shows a schematic example illustrating how the spatial distribution of member galaxies can lead to multiple candidate rotation axes. In this illustration, two possible orientations, marked as $\theta^1_{\rm max}$ and $\theta^2_{\rm max}$, both correspond to the same value of $\Delta Z_{\rm max}$. This highlights how, in cases of sparse or asymmetric galaxy distributions, it can be ambiguous to uniquely determine the projected rotation axis.

\begin{figure*}[!htp]
\plottwo{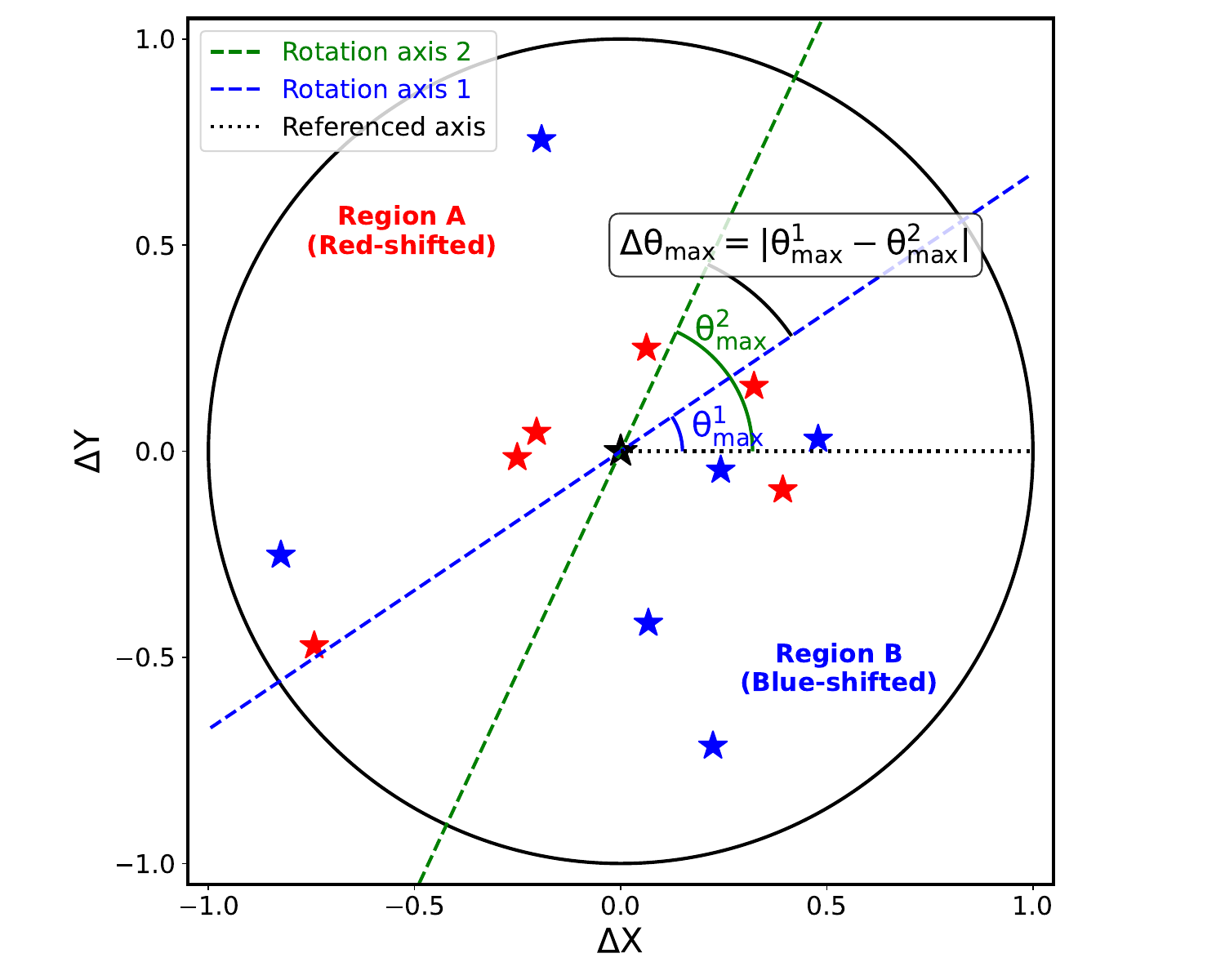}{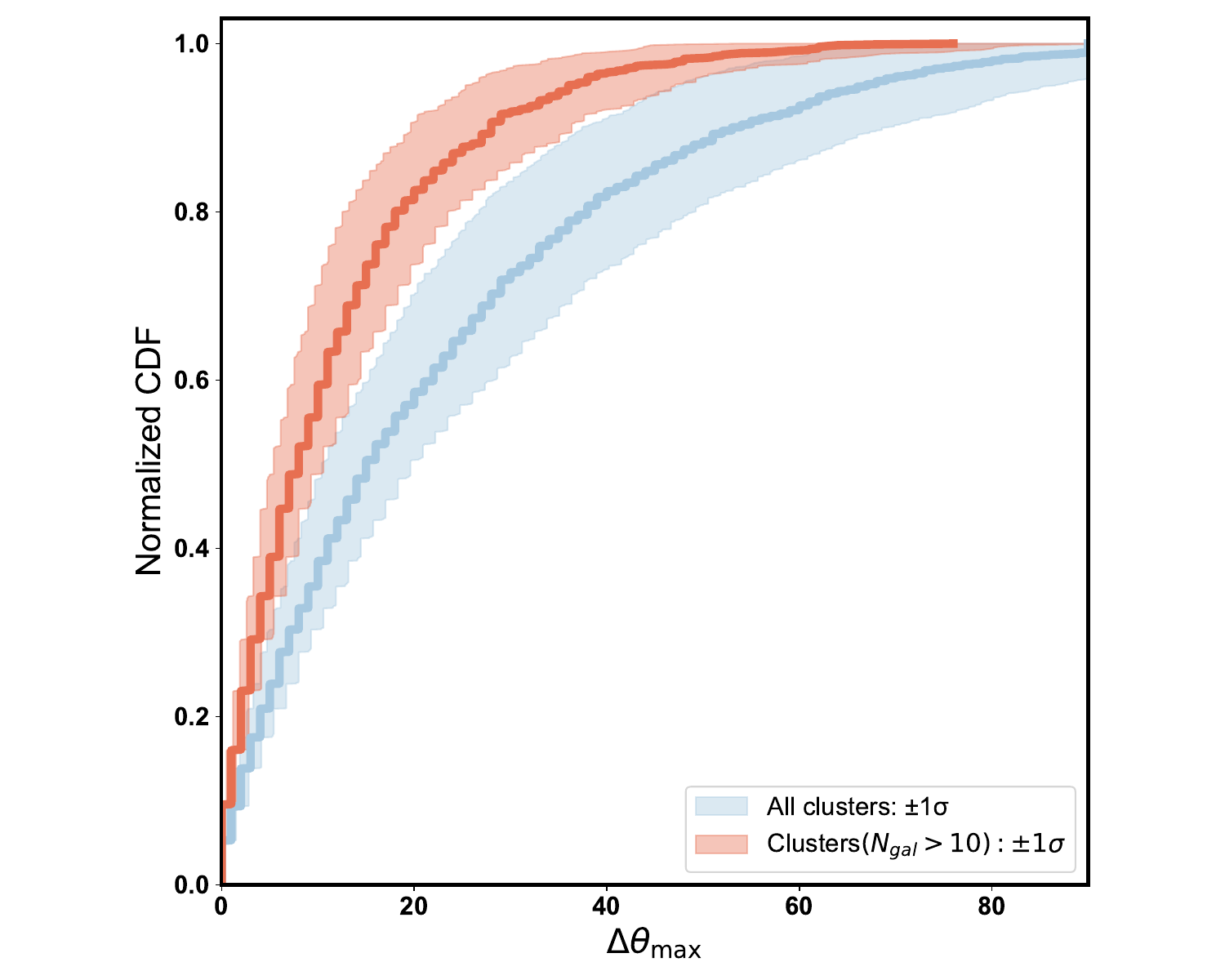}
\caption{\textbf{Left:} An example illustrating the multiplicity of $\theta_{\rm max}$ for a given $\Delta Z_{\rm max}$, due to the sparse and asymmetric distribution of member galaxies.
\textbf{Right:} The normalized cumulative distribution function (CDF) of the $\theta_{\rm max}$ for all clusters and clusters with richness larger than 10.}
\label{fig:appendix_theta_max}
\end{figure*}

The right panel of Figure~\ref{fig:appendix_theta_max} presents the normalized cumulative distribution function (CDF) of $\Delta \theta_{\rm max}$ for the entire cluster sample (blue) and for clusters with richness greater than 10 (orange). For the full sample, approximately 55\% of clusters have $\Delta \theta_{\rm max}$ less than $20^\circ$, indicating that in many cases the rotation axis is relatively well constrained. For the richer subsample with more than 10 member galaxies, this fraction increases to about 85\%, demonstrating that the ambiguity in determining $\Delta \theta_{\rm max}$ is significantly reduced when more member galaxies are available. This highlights the importance of sample richness in reliably identifying the projected rotation axis and minimizing uncertainties due to sparse or asymmetric galaxy distributions.

We recommend that, in cases where multiple candidate $\theta_{\rm max}$ values are found, additional criteria or statistical methods be considered to select the most physically meaningful rotation axis, or to quantify the associated uncertainty. Such approaches will help ensure the robustness of spin measurements, especially for clusters with limited spectroscopic data.

\section{Distributions of N and $\Delta Z_{\rm max}$, and their relationship}
\label{app:sec2}

This appendix provides additional summaries of cluster richness and the redshift contrast metric to complement the main text. Figure~A2 presents the distribution of the number of member galaxies per cluster ($N_{gal}$, left panel), the distribution of $\Delta Z_{\max}$ ( middle panel), and $\Delta Z_{\max}$ as a function of $N_{gal}$ (right panel) for both cluster samples. The sample-1 is dominated by sparse systems, with the majority of clusters containing fewer than 20 member galaxies,while Sample-2 includes clusters with a larger number of member galaxies. The distribution of $\Delta Z_{\max}$ is unimodal and peaks near 0.003 for Sample-1, whereas for Sample-2 it is smaller, peaking around 0.002. Consistent with the visual impression in the right panel, $\Delta Z_{\max}$ does not exhibit a strong dependence on cluster richness ($N_{\mathrm{gal}}$), although richer clusters show slightly larger scatter, and for Sample-2 there is a weak increase of $\Delta Z_{\max}$ with $N_{\mathrm{gal}}$.

\begin{figure*}[!htp]
\plotthree{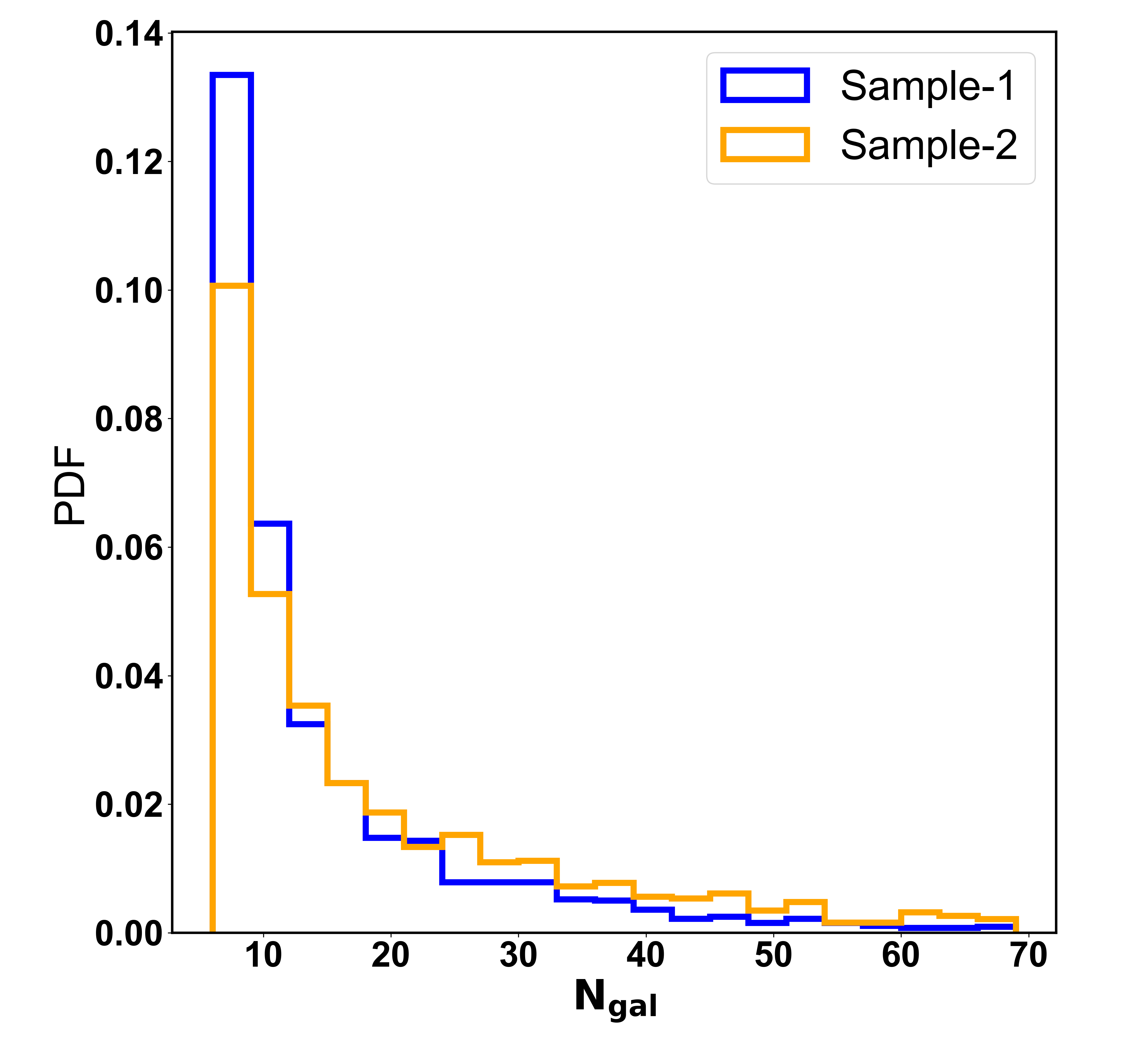}{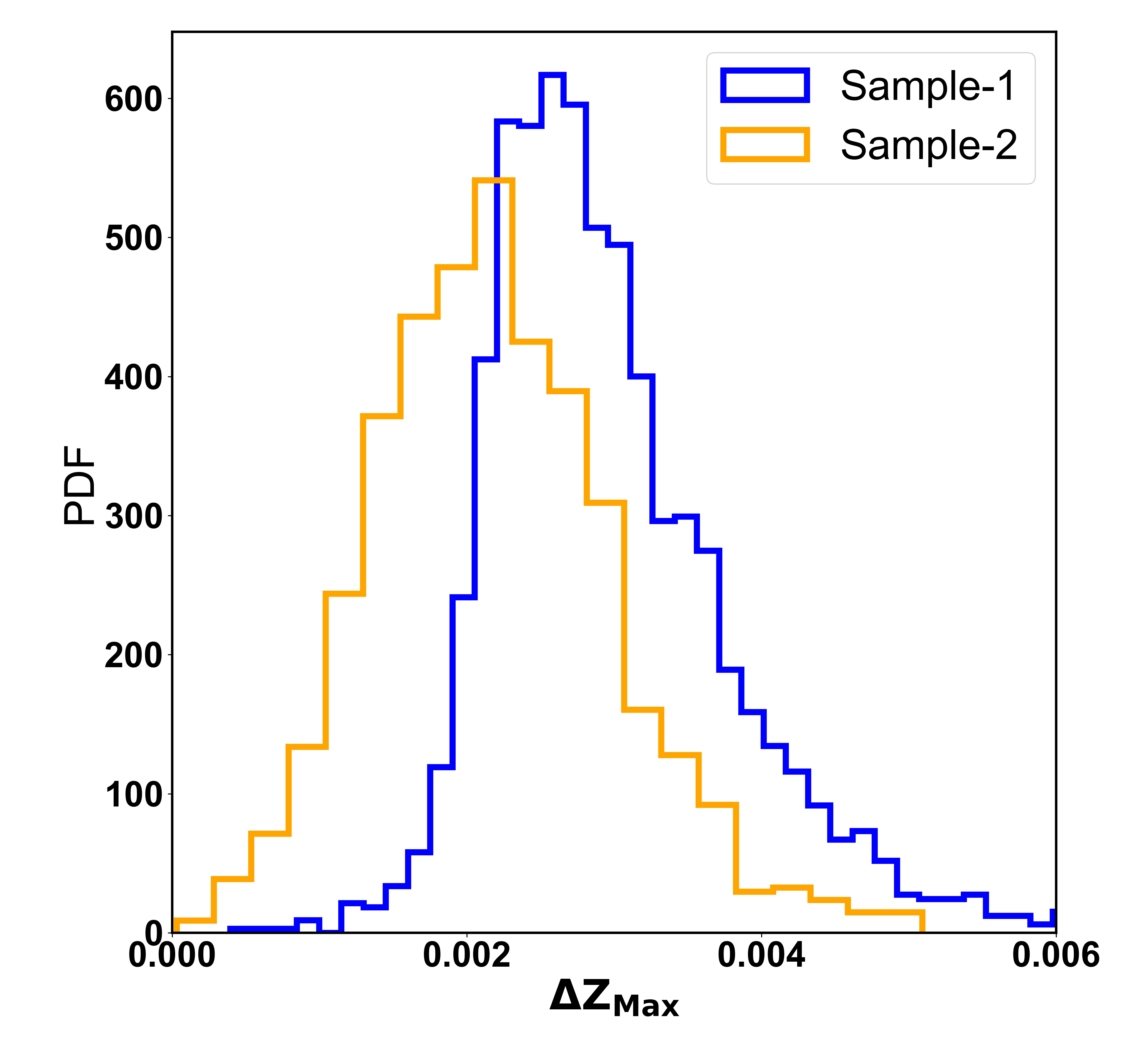}{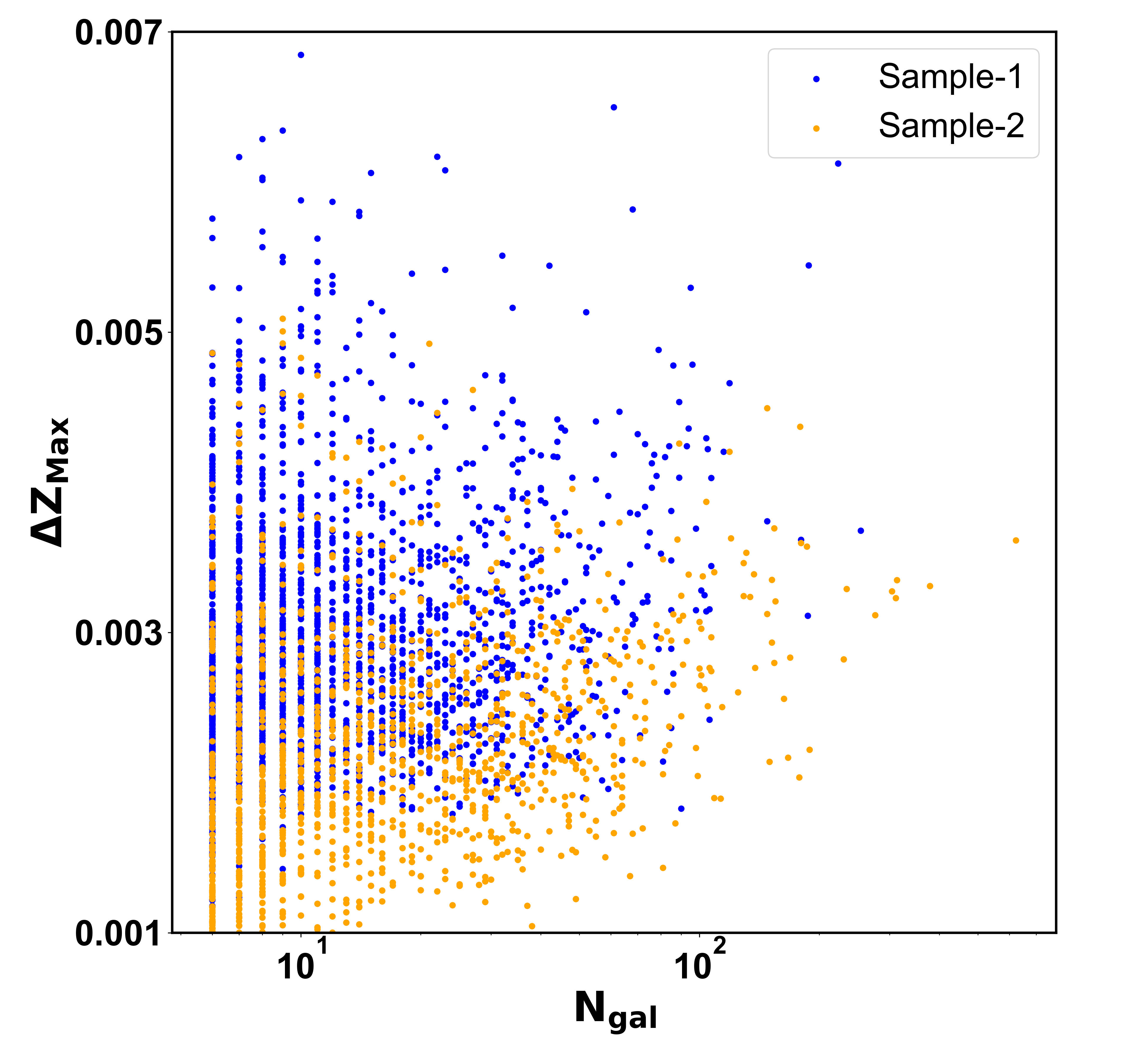}
\caption{Distributions of the basic properties of the galaxy cluster samples used for both Sample-1 and Sample-2. Left: distribution of the number of galaxies per cluster. Middle: distribution of $\Delta Z_{\max}$. Right: $\Delta Z_{\max}$ versus the number of galaxies per cluster.}
\label{fig:appendix_2}
\end{figure*}

\bibliography{main}{}

\begin{thebibliography}{}
\expandafter\ifx\csname natexlab\endcsname\relax\def\natexlab#1{#1}\fi

\bibitem[{{Abazajian} {et~al.}(2009){Abazajian}, {Adelman-McCarthy},
  {Ag{\"u}eros}, {Allam}, {Allende Prieto}, {An}, {Anderson}, {Anderson},
  {Annis}, {Bahcall}, {Bailer-Jones}, {Barentine}, {Bassett}, {Becker},
  {Beers}, {Bell}, {Belokurov}, {Berlind}, {Berman}, {Bernardi}, {Bickerton},
  {Bizyaev}, {Blakeslee}, {Blanton}, {Bochanski}, {Boroski}, {Brewington},
  {Brinchmann}, {Brinkmann}, {Brunner}, {Budav{\'a}ri}, {Carey}, {Carliles},
  {Carr}, {Castander}, {Cinabro}, {Connolly}, {Csabai}, {Cunha}, {Czarapata},
  {Davenport}, {de Haas}, {Dilday}, {Doi}, {Eisenstein}, {Evans}, {Evans},
  {Fan}, {Friedman}, {Frieman}, {Fukugita}, {G{\"a}nsicke}, {Gates},
  {Gillespie}, {Gilmore}, {Gonzalez}, {Gonzalez}, {Grebel}, {Gunn},
  {Gy{\"o}ry}, {Hall}, {Harding}, {Harris}, {Harvanek}, {Hawley}, {Hayes},
  {Heckman}, {Hendry}, {Hennessy}, {Hindsley}, {Hoblitt}, {Hogan}, {Hogg},
  {Holtzman}, {Hyde}, {Ichikawa}, {Ichikawa}, {Im}, {Ivezi{\'c}}, {Jester},
  {Jiang}, {Johnson}, {Jorgensen}, {Juri{\'c}}, {Kent}, {Kessler}, {Kleinman},
  {Knapp}, {Konishi}, {Kron}, {Krzesinski}, {Kuropatkin}, {Lampeitl},
  {Lebedeva}, {Lee}, {Lee}, {French Leger}, {L{\'e}pine}, {Li}, {Lima}, {Lin},
  {Long}, {Loomis}, {Loveday}, {Lupton}, {Magnier}, {Malanushenko},
  {Malanushenko}, {Mandelbaum}, {Margon}, {Marriner}, {Mart{\'\i}nez-Delgado},
  {Matsubara}, {McGehee}, {McKay}, {Meiksin}, {Morrison}, {Mullally}, {Munn},
  {Murphy}, {Nash}, {Nebot}, {Neilsen}, {Newberg}, {Newman}, {Nichol},
  {Nicinski}, {Nieto-Santisteban}, {Nitta}, {Okamura}, {Oravetz}, {Ostriker},
  {Owen}, {Padmanabhan}, {Pan}, {Park}, {Pauls}, {Peoples}, {Percival}, {Pier},
  {Pope}, {Pourbaix}, {Price}, {Purger}, {Quinn}, {Raddick}, {Re Fiorentin},
  {Richards}, {Richmond}, {Riess}, {Rix}, {Rockosi}, {Sako}, {Schlegel},
  {Schneider}, {Scholz}, {Schreiber}, {Schwope}, {Seljak}, {Sesar}, {Sheldon},
  {Shimasaku}, {Sibley}, {Simmons}, {Sivarani}, {Allyn Smith}, {Smith},
  {Smol{\v{c}}i{\'c}}, {Snedden}, {Stebbins}, {Steinmetz}, {Stoughton},
  {Strauss}, {SubbaRao}, {Suto}, {Szalay}, {Szapudi}, {Szkody}, {Tanaka},
  {Tegmark}, {Teodoro}, {Thakar}, {Tremonti}, {Tucker}, {Uomoto}, {Vanden
  Berk}, {Vandenberg}, {Vidrih}, {Vogeley}, {Voges}, {Vogt}, {Wadadekar},
  {Watters}, {Weinberg}, {West}, {White}, {Wilhite}, {Wonders}, {Yanny}, \&
  {Yocum}}]{2009ApJS..182..543A}
{Abazajian}, K.~N., {Adelman-McCarthy}, J.~K., {Ag{\"u}eros}, M.~A., {et~al.}
  2009, \apjs, 182, 543

\bibitem[{{Alam} {et~al.}(2015){Alam}, {Albareti}, {Allende Prieto}, {Anders},
  {Anderson}, {Anderton}, {Andrews}, {Armengaud}, {Aubourg}, {Bailey}, {Basu},
  {Bautista}, {Beaton}, {Beers}, {Bender}, {Berlind}, {Beutler}, {Bhardwaj},
  {Bird}, {Bizyaev}, {Blake}, {Blanton}, {Blomqvist}, {Bochanski}, {Bolton},
  {Bovy}, {Shelden Bradley}, {Brandt}, {Brauer}, {Brinkmann}, {Brown},
  {Brownstein}, {Burden}, {Burtin}, {Busca}, {Cai}, {Capozzi}, {Carnero
  Rosell}, {Carr}, {Carrera}, {Chambers}, {Chaplin}, {Chen}, {Chiappini},
  {Chojnowski}, {Chuang}, {Clerc}, {Comparat}, {Covey}, {Croft}, {Cuesta},
  {Cunha}, {da Costa}, {Da Rio}, {Davenport}, {Dawson}, {De Lee}, {Delubac},
  {Deshpande}, {Dhital}, {Dutra-Ferreira}, {Dwelly}, {Ealet}, {Ebelke},
  {Edmondson}, {Eisenstein}, {Ellsworth}, {Elsworth}, {Epstein}, {Eracleous},
  {Escoffier}, {Esposito}, {Evans}, {Fan}, {Fern{\'a}ndez-Alvar}, {Feuillet},
  {Filiz Ak}, {Finley}, {Finoguenov}, {Flaherty}, {Fleming}, {Font-Ribera},
  {Foster}, {Frinchaboy}, {Galbraith-Frew}, {Garc{\'\i}a},
  {Garc{\'\i}a-Hern{\'a}ndez}, {Garc{\'\i}a P{\'e}rez}, {Gaulme}, {Ge},
  {G{\'e}nova-Santos}, {Georgakakis}, {Ghezzi}, {Gillespie}, {Girardi},
  {Goddard}, {Gontcho}, {Gonz{\'a}lez Hern{\'a}ndez}, {Grebel}, {Green},
  {Grieb}, {Grieves}, {Gunn}, {Guo}, {Harding}, {Hasselquist}, {Hawley},
  {Hayden}, {Hearty}, {Hekker}, {Ho}, {Hogg}, {Holley-Bockelmann}, {Holtzman},
  {Honscheid}, {Huber}, {Huehnerhoff}, {Ivans}, {Jiang}, {Johnson},
  {Kinemuchi}, {Kirkby}, {Kitaura}, {Klaene}, {Knapp}, {Kneib}, {Koenig},
  {Lam}, {Lan}, {Lang}, {Laurent}, {Le Goff}, {Leauthaud}, {Lee}, {Lee},
  {Licquia}, {Liu}, {Long}, {L{\'o}pez-Corredoira}, {Lorenzo-Oliveira},
  {Lucatello}, {Lundgren}, {Lupton}, {Mack}, {Mahadevan}, {Maia}, {Majewski},
  {Malanushenko}, {Malanushenko}, {Manchado}, {Manera}, {Mao}, {Maraston},
  {Marchwinski}, {Margala}, {Martell}, {Martig}, {Masters}, {Mathur},
  {McBride}, {McGehee}, {McGreer}, {McMahon}, {M{\'e}nard}, {Menzel},
  {Merloni}, {M{\'e}sz{\'a}ros}, {Miller}, {Miralda-Escud{\'e}}, {Miyatake},
  {Montero-Dorta}, {More}, {Morganson}, {Morice-Atkinson}, {Morrison},
  {Mosser}, {Muna}, {Myers}, {Nandra}, {Newman}, {Neyrinck}, {Nguyen},
  {Nichol}, {Nidever}, {Noterdaeme}, {Nuza}, {O'Connell}, {O'Connell},
  {O'Connell}, {Ogando}, {Olmstead}, {Oravetz}, {Oravetz}, {Osumi}, {Owen},
  {Padgett}, {Padmanabhan}, {Paegert}, {Palanque-Delabrouille}, \&
  {Pan}}]{2015ApJS..219...12A}
{Alam}, S., {Albareti}, F.~D., {Allende Prieto}, C., {et~al.} 2015, \apjs, 219,
  12

\bibitem[{{Altamura} {et~al.}(2023){Altamura}, {Kay}, {Chluba}, \&
  {Towler}}]{2023MNRAS.524.2262A}
{Altamura}, E., {Kay}, S.~T., {Chluba}, J., \& {Towler}, I. 2023, \mnras, 524,
  2262

\bibitem[{{Baldi} {et~al.}(2018){Baldi}, {De Petris}, {Sembolini}, {Yepes},
  {Cui}, \& {Lamagna}}]{2018MNRAS.479.4028B}
{Baldi}, A.~S., {De Petris}, M., {Sembolini}, F., {et~al.} 2018, \mnras, 479,
  4028

\bibitem[{{Baldi} {et~al.}(2019){Baldi}, {De Petris}, {Sembolini}, {Yepes},
  {Cui}, \& {Lamagna}}]{2019JPhCS1226a2003B}
{Baldi}, A.~S., {De Petris}, M., {Sembolini}, F., {et~al.} 2019, in Journal of
  Physics Conference Series, Vol. 1226, Journal of Physics Conference Series
  (IOP), 012003

\bibitem[{{Baldi} {et~al.}(2017){Baldi}, {De Petris}, {Sembolini}, {Yepes},
  {Lamagna}, \& {Rasia}}]{2017MNRAS.465.2584B}
---. 2017, \mnras, 465, 2584

\bibitem[{{Barnes} {et~al.}(1985){Barnes}, {Dekel}, {Efstathiou}, \&
  {Frenk}}]{1985ApJ...295..368B}
{Barnes}, J., {Dekel}, A., {Efstathiou}, G., \& {Frenk}, C.~S. 1985, \apj, 295,
  368

\bibitem[{{Barreira} {et~al.}(2015){Barreira}, {Li}, {Jennings}, {Merten},
  {King}, {Baugh}, \& {Pascoli}}]{2015MNRAS.454.4085B}
{Barreira}, A., {Li}, B., {Jennings}, E., {et~al.} 2015, \mnras, 454, 4085

\bibitem[{{Berlind} {et~al.}(2006){Berlind}, {Frieman}, {Weinberg}, {Blanton},
  {Warren}, {Abazajian}, {Scranton}, {Hogg}, {Scoccimarro}, {Bahcall},
  {Brinkmann}, {Gott}, {Kleinman}, {Krzesinski}, {Lee}, {Miller}, {Nitta},
  {Schneider}, {Tucker}, {Zehavi}, \& {SDSS
  Collaboration}}]{2006ApJS..167....1B}
{Berlind}, A.~A., {Frieman}, J., {Weinberg}, D.~H., {et~al.} 2006, \apjs, 167,
  1

\bibitem[{{Bilton} {et~al.}(2019){Bilton}, {Hunt}, {Pimbblet}, \&
  {Roediger}}]{2019MNRAS.490.5017B}
{Bilton}, L.~E., {Hunt}, M., {Pimbblet}, K.~A., \& {Roediger}, E. 2019, \mnras,
  490, 5017

\bibitem[{{Binggeli}(1982)}]{1982A&A...107..338B}
{Binggeli}, B. 1982, \aap, 107, 338

\bibitem[{{Blanton} {et~al.}(2005){Blanton}, {Schlegel}, {Strauss},
  {Brinkmann}, {Finkbeiner}, {Fukugita}, {Gunn}, {Hogg}, {Ivezi{\'c}}, {Knapp},
  {Lupton}, {Munn}, {Schneider}, {Tegmark}, \& {Zehavi}}]{2005AJ....129.2562B}
{Blanton}, M.~R., {Schlegel}, D.~J., {Strauss}, M.~A., {et~al.} 2005, \aj, 129,
  2562

\bibitem[{{Burgett} {et~al.}(2004){Burgett}, {Vick}, {Davis}, {Colless}, {De
  Propris}, {Baldry}, {Baugh}, {Bland-Hawthorn}, {Bridges}, {Cannon}, {Cole},
  {Collins}, {Couch}, {Cross}, {Dalton}, {Driver}, {Efstathiou}, {Ellis},
  {Frenk}, {Glazebrook}, {Hawkins}, {Jackson}, {Lahav}, {Lewis}, {Lumsden},
  {Maddox}, {Madgwick}, {Norberg}, {Peacock}, {Percival}, {Peterson},
  {Sutherland}, \& {Taylor}}]{2004MNRAS.352..605B}
{Burgett}, W.~S., {Vick}, M.~M., {Davis}, D.~S., {et~al.} 2004, \mnras, 352,
  605

\bibitem[{{Cautun} {et~al.}(2014){Cautun}, {van de Weygaert}, {Jones}, \&
  {Frenk}}]{2014MNRAS.441.2923C}
{Cautun}, M., {van de Weygaert}, R., {Jones}, B. J.~T., \& {Frenk}, C.~S. 2014,
  \mnras, 441, 2923

\bibitem[{{Chluba} \& {Mannheim}(2002)}]{2002A&A...396..419C}
{Chluba}, J., \& {Mannheim}, K. 2002, \aap, 396, 419

\bibitem[{{Codis} {et~al.}(2012){Codis}, {Pichon}, {Devriendt}, {Slyz},
  {Pogosyan}, {Dubois}, \& {Sousbie}}]{2012MNRAS.427.3320C}
{Codis}, S., {Pichon}, C., {Devriendt}, J., {et~al.} 2012, \mnras, 427, 3320

\bibitem[{{den Hartog} \& {Katgert}(1996)}]{1996MNRAS.279..349D}
{den Hartog}, R., \& {Katgert}, P. 1996, \mnras, 279, 349

\bibitem[{{Dolag} {et~al.}(2009){Dolag}, {Borgani}, {Murante}, \&
  {Springel}}]{2009MNRAS.399..497D}
{Dolag}, K., {Borgani}, S., {Murante}, G., \& {Springel}, V. 2009, \mnras, 399,
  497

\bibitem[{{Fox} {et~al.}(2022){Fox}, {Mahler}, {Sharon}, \& {Remolina
  Gonz{\'a}lez}}]{2022ApJ...928...87F}
{Fox}, C., {Mahler}, G., {Sharon}, K., \& {Remolina Gonz{\'a}lez}, J.~D. 2022,
  \apj, 928, 87

\bibitem[{{Ganeshaiah Veena} {et~al.}(2019){Ganeshaiah Veena}, {Cautun},
  {Tempel}, {van de Weygaert}, \& {Frenk}}]{2019MNRAS.487.1607G}
{Ganeshaiah Veena}, P., {Cautun}, M., {Tempel}, E., {van de Weygaert}, R., \&
  {Frenk}, C.~S. 2019, \mnras, 487, 1607

\bibitem[{{Ganeshaiah Veena} {et~al.}(2021){Ganeshaiah Veena}, {Cautun}, {van
  de Weygaert}, {Tempel}, \& {Frenk}}]{2021MNRAS.503.2280G}
{Ganeshaiah Veena}, P., {Cautun}, M., {van de Weygaert}, R., {Tempel}, E., \&
  {Frenk}, C.~S. 2021, \mnras, 503, 2280

\bibitem[{{Ganeshaiah Veena} {et~al.}(2018){Ganeshaiah Veena}, {Cautun}, {van
  de Weygaert}, {Tempel}, {Jones}, {Rieder}, \& {Frenk}}]{2018MNRAS.481..414G}
{Ganeshaiah Veena}, P., {Cautun}, M., {van de Weygaert}, R., {et~al.} 2018,
  \mnras, 481, 414

\bibitem[{{Graham} \& {Cappellari}(2023)}]{2023A&A...675A.161G}
{Graham}, M.~T., \& {Cappellari}, M. 2023, \aap, 675, A161

\bibitem[{{Hamden} {et~al.}(2010){Hamden}, {Simpson}, {Johnston}, \&
  {Lee}}]{2010ApJ...716L.205H}
{Hamden}, E.~T., {Simpson}, C.~M., {Johnston}, K.~V., \& {Lee}, D.~M. 2010,
  \apjl, 716, L205

\bibitem[{{Huang} {et~al.}(2016){Huang}, {Mandelbaum}, {Freeman}, {Chen},
  {Rozo}, {Rykoff}, \& {Baxter}}]{2016MNRAS.463..222H}
{Huang}, H.-J., {Mandelbaum}, R., {Freeman}, P.~E., {et~al.} 2016, \mnras, 463,
  222

\bibitem[{{Huchra} \& {Geller}(1982)}]{1982ApJ...257..423H}
{Huchra}, J.~P., \& {Geller}, M.~J. 1982, \apj, 257, 423

\bibitem[{{Hwang} \& {Lee}(2007)}]{2007ApJ...662..236H}
{Hwang}, H.~S., \& {Lee}, M.~G. 2007, \apj, 662, 236

\bibitem[{{Kalinkov} {et~al.}(2005){Kalinkov}, {Valchanov}, {Valtchanov},
  {Kuneva}, \& {Dissanska}}]{2005MNRAS.359.1491K}
{Kalinkov}, M., {Valchanov}, T., {Valtchanov}, I., {Kuneva}, I., \&
  {Dissanska}, M. 2005, \mnras, 359, 1491

\bibitem[{{Kang} \& {Wang}(2015)}]{2015ApJ...813....6K}
{Kang}, X., \& {Wang}, P. 2015, \apj, 813, 6

\bibitem[{{Kraljic} {et~al.}(2021){Kraljic}, {Duckworth}, {Tojeiro}, {Alam},
  {Bizyaev}, {Weijmans}, {Boardman}, \& {Lane}}]{2021MNRAS.504.4626K}
{Kraljic}, K., {Duckworth}, C., {Tojeiro}, R., {et~al.} 2021, \mnras, 504, 4626

\bibitem[{{Kugel} \& {van de Weygaert}(2024)}]{2024arXiv240716489K}
{Kugel}, R., \& {van de Weygaert}, R. 2024, arXiv e-prints, arXiv:2407.16489

\bibitem[{{Laigle} {et~al.}(2015){Laigle}, {Pichon}, {Codis}, {Dubois}, {Le
  Borgne}, {Pogosyan}, {Devriendt}, {Peirani}, {Prunet}, {Rouberol}, {Slyz}, \&
  {Sousbie}}]{2015MNRAS.446.2744L}
{Laigle}, C., {Pichon}, C., {Codis}, S., {et~al.} 2015, \mnras, 446, 2744

\bibitem[{{Lee} \& {Erdogdu}(2007)}]{2007ApJ...671.1248L}
{Lee}, J., \& {Erdogdu}, P. 2007, \apj, 671, 1248

\bibitem[{{Libeskind} {et~al.}(2013){Libeskind}, {Hoffman}, {Forero-Romero},
  {Gottl{\"o}ber}, {Knebe}, {Steinmetz}, \& {Klypin}}]{2013MNRAS.428.2489L}
{Libeskind}, N.~I., {Hoffman}, Y., {Forero-Romero}, J., {et~al.} 2013, \mnras,
  428, 2489

\bibitem[{{Libeskind} {et~al.}(2014){Libeskind}, {Knebe}, {Hoffman}, \&
  {Gottl{\"o}ber}}]{2014MNRAS.443.1274L}
{Libeskind}, N.~I., {Knebe}, A., {Hoffman}, Y., \& {Gottl{\"o}ber}, S. 2014,
  \mnras, 443, 1274

\bibitem[{{Libeskind} {et~al.}(2018){Libeskind}, {van de Weygaert}, {Cautun},
  {Falck}, {Tempel}, {Abel}, {Alpaslan}, {Arag{\'o}n-Calvo}, {Forero-Romero},
  {Gonzalez}, {Gottl{\"o}ber}, {Hahn}, {Hellwing}, {Hoffman}, {Jones},
  {Kitaura}, {Knebe}, {Manti}, {Neyrinck}, {Nuza}, {Padilla}, {Platen},
  {Ramachandra}, {Robotham}, {Saar}, {Shandarin}, {Steinmetz}, {Stoica},
  {Sousbie}, \& {Yepes}}]{2018MNRAS.473.1195L}
{Libeskind}, N.~I., {van de Weygaert}, R., {Cautun}, M., {et~al.} 2018, \mnras,
  473, 1195

\bibitem[{{L{\'o}pez}(2024)}]{2024PASP..136c7001L}
{L{\'o}pez}, P. 2024, \pasp, 136, 037001

\bibitem[{{L{\'o}pez} {et~al.}(2021){L{\'o}pez}, {Cautun}, {Paz},
  {Merch{\'a}n}, \& {van de Weygaert}}]{2021MNRAS.502.5528L}
{L{\'o}pez}, P., {Cautun}, M., {Paz}, D., {Merch{\'a}n}, M., \& {van de
  Weygaert}, R. 2021, \mnras, 502, 5528

\bibitem[{{L{\'o}pez} {et~al.}(2025){L{\'o}pez}, {van de Weygaert}, \&
  {Merch{\'a}n}}]{2025arXiv250501298L}
{L{\'o}pez}, P., {van de Weygaert}, R., \& {Merch{\'a}n}, M. 2025, arXiv
  e-prints, arXiv:2505.01298

\bibitem[{{Manolopoulou} \& {Plionis}(2017)}]{2017MNRAS.465.2616M}
{Manolopoulou}, M., \& {Plionis}, M. 2017, \mnras, 465, 2616

\bibitem[{{Materne} \& {Hopp}(1983)}]{1983A&A...124L..13M}
{Materne}, J., \& {Hopp}, U. 1983, \aap, 124, L13

\bibitem[{{McNamara}(2012)}]{2012gcgc.conf...37M}
{McNamara}, B. 2012, in Galaxy Clusters as Giant Cosmic Laboratories, ed. J.-U.
  {Ness}, 37

\bibitem[{{Miller} {et~al.}(2005){Miller}, {Nichol}, {Reichart}, {Wechsler},
  {Evrard}, {Annis}, {McKay}, {Bahcall}, {Bernardi}, {Boehringer}, {Connolly},
  {Goto}, {Kniazev}, {Lamb}, {Postman}, {Schneider}, {Sheth}, \&
  {Voges}}]{2005AJ....130..968M}
{Miller}, C.~J., {Nichol}, R.~C., {Reichart}, D., {et~al.} 2005, \aj, 130, 968

\bibitem[{{Nelson} {et~al.}(2019){Nelson}, {Springel}, {Pillepich},
  {Rodriguez-Gomez}, {Torrey}, {Genel}, {Vogelsberger}, {Pakmor}, {Marinacci},
  {Weinberger}, {Kelley}, {Lovell}, {Diemer}, \&
  {Hernquist}}]{2019ComAC...6....2N}
{Nelson}, D., {Springel}, V., {Pillepich}, A., {et~al.} 2019, Computational
  Astrophysics and Cosmology, 6, 2

\bibitem[{{Niederste-Ostholt} {et~al.}(2010){Niederste-Ostholt}, {Strauss},
  {Dong}, {Koester}, \& {McKay}}]{2010MNRAS.405.2023N}
{Niederste-Ostholt}, M., {Strauss}, M.~A., {Dong}, F., {Koester}, B.~P., \&
  {McKay}, T.~A. 2010, \mnras, 405, 2023

\bibitem[{{Oegerle} \& {Hill}(1992)}]{1992AJ....104.2078O}
{Oegerle}, W.~R., \& {Hill}, J.~M. 1992, \aj, 104, 2078

\bibitem[{{Peebles}(1969)}]{1969ApJ...155..393P}
{Peebles}, P.~J.~E. 1969, \apj, 155, 393

\bibitem[{{Pillepich} {et~al.}(2018){Pillepich}, {Nelson}, {Hernquist},
  {Springel}, {Pakmor}, {Torrey}, {Weinberger}, {Genel}, {Naiman}, {Marinacci},
  \& {Vogelsberger}}]{2018MNRAS.475..648P}
{Pillepich}, A., {Nelson}, D., {Hernquist}, L., {et~al.} 2018, \mnras, 475, 648

\bibitem[{{Planck Collaboration} {et~al.}(2016){Planck Collaboration}, {Ade},
  {Aghanim}, {Arnaud}, {Ashdown}, {Aumont}, {Baccigalupi}, {Banday},
  {Barreiro}, {Bartlett}, {Bartolo}, {Battaner}, {Battye}, {Benabed},
  {Beno{\^\i}t}, {Benoit-L{\'e}vy}, {Bernard}, {Bersanelli}, {Bielewicz},
  {Bock}, {Bonaldi}, {Bonavera}, {Bond}, {Borrill}, {Bouchet}, {Boulanger},
  {Bucher}, {Burigana}, {Butler}, {Calabrese}, {Cardoso}, {Catalano},
  {Challinor}, {Chamballu}, {Chary}, {Chiang}, {Chluba}, {Christensen},
  {Church}, {Clements}, {Colombi}, {Colombo}, {Combet}, {Coulais}, {Crill},
  {Curto}, {Cuttaia}, {Danese}, {Davies}, {Davis}, {de Bernardis}, {de Rosa},
  {de Zotti}, {Delabrouille}, {D{\'e}sert}, {Di Valentino}, {Dickinson},
  {Diego}, {Dolag}, {Dole}, {Donzelli}, {Dor{\'e}}, {Douspis}, {Ducout},
  {Dunkley}, {Dupac}, {Efstathiou}, {Elsner}, {En{\ss}lin}, {Eriksen},
  {Farhang}, {Fergusson}, {Finelli}, {Forni}, {Frailis}, {Fraisse},
  {Franceschi}, {Frejsel}, {Galeotta}, {Galli}, {Ganga}, {Gauthier}, {Gerbino},
  {Ghosh}, {Giard}, {Giraud-H{\'e}raud}, {Giusarma}, {Gjerl{\o}w},
  {Gonz{\'a}lez-Nuevo}, {G{\'o}rski}, {Gratton}, {Gregorio}, {Gruppuso},
  {Gudmundsson}, {Hamann}, {Hansen}, {Hanson}, {Harrison}, {Helou},
  {Henrot-Versill{\'e}}, {Hern{\'a}ndez-Monteagudo}, {Herranz}, {Hildebrandt},
  {Hivon}, {Hobson}, {Holmes}, {Hornstrup}, {Hovest}, {Huang}, {Huffenberger},
  {Hurier}, {Jaffe}, {Jaffe}, {Jones}, {Juvela}, {Keih{\"a}nen}, {Keskitalo},
  {Kisner}, {Kneissl}, {Knoche}, {Knox}, {Kunz}, {Kurki-Suonio}, {Lagache},
  {L{\"a}hteenm{\"a}ki}, {Lamarre}, {Lasenby}, {Lattanzi}, {Lawrence}, {Leahy},
  {Leonardi}, {Lesgourgues}, {Levrier}, {Lewis}, {Liguori}, {Lilje},
  {Linden-V{\o}rnle}, {L{\'o}pez-Caniego}, {Lubin}, {Mac{\'\i}as-P{\'e}rez},
  {Maggio}, {Maino}, {Mandolesi}, {Mangilli}, {Marchini}, {Maris}, {Martin},
  {Martinelli}, {Mart{\'\i}nez-Gonz{\'a}lez}, {Masi}, {Matarrese}, {McGehee},
  {Meinhold}, {Melchiorri}, {Melin}, {Mendes}, {Mennella}, {Migliaccio},
  {Millea}, {Mitra}, {Miville-Desch{\^e}nes}, {Moneti}, {Montier}, {Morgante},
  {Mortlock}, {Moss}, {Munshi}, {Murphy}, {Naselsky}, {Nati}, {Natoli},
  {Netterfield}, {N{\o}rgaard-Nielsen}, {Noviello}, {Novikov}, {Novikov},
  {Oxborrow}, {Paci}, {Pagano}, {Pajot}, {Paladini}, {Paoletti}, {Partridge},
  {Pasian}, {Patanchon}, {Pearson}, {Perdereau}, {Perotto}, {Perrotta},
  {Pettorino}, {Piacentini}, {Piat}, {Pierpaoli}, {Pietrobon}, {Plaszczynski},
  {Pointecouteau}, {Polenta}, {Popa}, {Pratt}, \&
  {Pr{\'e}zeau}}]{2016A&A...594A..13P}
{Planck Collaboration}, {Ade}, P.~A.~R., {Aghanim}, N., {et~al.} 2016, \aap,
  594, A13

\bibitem[{{Porciani} {et~al.}(2002{\natexlab{a}}){Porciani}, {Dekel}, \&
  {Hoffman}}]{2002MNRAS.332..325P}
{Porciani}, C., {Dekel}, A., \& {Hoffman}, Y. 2002{\natexlab{a}}, \mnras, 332,
  325

\bibitem[{{Porciani} {et~al.}(2002{\natexlab{b}}){Porciani}, {Dekel}, \&
  {Hoffman}}]{2002MNRAS.332..339P}
---. 2002{\natexlab{b}}, \mnras, 332, 339

\bibitem[{{Rong} {et~al.}(2025){Rong}, {Wang}, \& {Tang}}]{2025ApJ...983L...3R}
{Rong}, Y., {Wang}, P., \& {Tang}, X.-x. 2025, \apjl, 983, L3

\bibitem[{{Sastry}(1968)}]{1968PASP...80..252S}
{Sastry}, G.~N. 1968, \pasp, 80, 252

\bibitem[{{Sharon} {et~al.}(2020){Sharon}, {Bayliss}, {Dahle}, {Dunham},
  {Florian}, {Gladders}, {Johnson}, {Mahler}, {Paterno-Mahler}, {Rigby},
  {Whitaker}, {Akhshik}, {Koester}, {Murray}, {Remolina Gonz{\'a}lez}, \&
  {Wuyts}}]{2020ApJS..247...12S}
{Sharon}, K., {Bayliss}, M.~B., {Dahle}, H., {et~al.} 2020, \apjs, 247, 12

\bibitem[{{Springel} {et~al.}(2005){Springel}, {White}, {Jenkins}, {Frenk},
  {Yoshida}, {Gao}, {Navarro}, {Thacker}, {Croton}, {Helly}, {Peacock}, {Cole},
  {Thomas}, {Couchman}, {Evrard}, {Colberg}, \& {Pearce}}]{2005Natur.435..629S}
{Springel}, V., {White}, S. D.~M., {Jenkins}, A., {et~al.} 2005, \nat, 435, 629

\bibitem[{{Springel} {et~al.}(2018){Springel}, {Pakmor}, {Pillepich},
  {Weinberger}, {Nelson}, {Hernquist}, {Vogelsberger}, {Genel}, {Torrey},
  {Marinacci}, \& {Naiman}}]{2018MNRAS.475..676S}
{Springel}, V., {Pakmor}, R., {Pillepich}, A., {et~al.} 2018, \mnras, 475, 676

\bibitem[{{Tang} {et~al.}(2025){Tang}, {Wang}, {Wang}, {Sheng}, {Yu}, \&
  {Xu}}]{2025ApJ...982..197T}
{Tang}, X.-x., {Wang}, P., {Wang}, W., {et~al.} 2025, \apj, 982, 197

\bibitem[{{Tempel} {et~al.}(2016){Tempel}, {Kipper}, {Tamm}, {Gramann},
  {Einasto}, {Sepp}, \& {Tuvikene}}]{2016A&A...588A..14T}
{Tempel}, E., {Kipper}, R., {Tamm}, A., {et~al.} 2016, \aap, 588, A14

\bibitem[{{Tempel} {et~al.}(2014{\natexlab{a}}){Tempel}, {Stoica},
  {Mart{\'\i}nez}, {Liivam{\"a}gi}, {Castellan}, \&
  {Saar}}]{2014MNRAS.438.3465T}
{Tempel}, E., {Stoica}, R.~S., {Mart{\'\i}nez}, V.~J., {et~al.}
  2014{\natexlab{a}}, \mnras, 438, 3465

\bibitem[{{Tempel} {et~al.}(2017){Tempel}, {Tuvikene}, {Kipper}, \&
  {Libeskind}}]{2017A&A...602A.100T}
{Tempel}, E., {Tuvikene}, T., {Kipper}, R., \& {Libeskind}, N.~I. 2017, \aap,
  602, A100

\bibitem[{{Tempel} {et~al.}(2014{\natexlab{b}}){Tempel}, {Tamm}, {Gramann},
  {Tuvikene}, {Liivam{\"a}gi}, {Suhhonenko}, {Kipper}, {Einasto}, \&
  {Saar}}]{2014A&A...566A...1T}
{Tempel}, E., {Tamm}, A., {Gramann}, M., {et~al.} 2014{\natexlab{b}}, \aap,
  566, A1

\bibitem[{{Umetsu}(2020)}]{2020A&ARv..28....7U}
{Umetsu}, K. 2020, \aapr, 28, 7

\bibitem[{{van de Weygaert} \& {Bond}(2008)}]{2008LNP...740..335V}
{van de Weygaert}, R., \& {Bond}, J.~R. 2008, in A Pan-Chromatic View of
  Clusters of Galaxies and the Large-Scale Structure, ed. M.~{Plionis},
  O.~{L{\'o}pez-Cruz}, \& D.~{Hughes}, Vol. 740, 335

\bibitem[{{Varela} {et~al.}(2012){Varela}, {Betancort-Rijo}, {Trujillo}, \&
  {Ricciardelli}}]{2012ApJ...744...82V}
{Varela}, J., {Betancort-Rijo}, J., {Trujillo}, I., \& {Ricciardelli}, E. 2012,
  \apj, 744, 82

\bibitem[{{Vitvitska} {et~al.}(2002){Vitvitska}, {Klypin}, {Kravtsov},
  {Wechsler}, {Primack}, \& {Bullock}}]{2002ApJ...581..799V}
{Vitvitska}, M., {Klypin}, A.~A., {Kravtsov}, A.~V., {et~al.} 2002, \apj, 581,
  799

\bibitem[{{Wake} {et~al.}(2017){Wake}, {Bundy}, {Diamond-Stanic}, {Yan},
  {Blanton}, {Bershady}, {S{\'a}nchez-Gallego}, {Drory}, {Jones}, {Kauffmann},
  {Law}, {Li}, {MacDonald}, {Masters}, {Thomas}, {Tinker}, {Weijmans}, \&
  {Brownstein}}]{2017AJ....154...86W}
{Wake}, D.~A., {Bundy}, K., {Diamond-Stanic}, A.~M., {et~al.} 2017, \aj, 154,
  86

\bibitem[{{Wang} {et~al.}(2025{\natexlab{a}}){Wang}, {Wang}, {Bao}, {Chen},
  {Tang}, {Zhang}, {Kang}, {Guo}, {Sheng}, \& {Yu}}]{2025ApJ...987L..30W}
{Wang}, H.-d., {Wang}, P., {Bao}, M., {et~al.} 2025{\natexlab{a}}, \apjl, 987,
  L30

\bibitem[{{Wang}(2025)}]{2025ApJ...992L..17W}
{Wang}, P. 2025, \apjl, 992, L17

\bibitem[{{Wang} \& {Kang}(2017)}]{2017MNRAS.468L.123W}
{Wang}, P., \& {Kang}, X. 2017, \mnras, 468, L123

\bibitem[{{Wang} \& {Kang}(2018)}]{2018MNRAS.473.1562W}
---. 2018, \mnras, 473, 1562

\bibitem[{{Wang} {et~al.}(2021){Wang}, {Libeskind}, {Tempel}, {Kang}, \&
  {Guo}}]{2021NatAs...5..839W}
{Wang}, P., {Libeskind}, N.~I., {Tempel}, E., {Kang}, X., \& {Guo}, Q. 2021,
  Nature Astronomy, 5, 839

\bibitem[{{Wang} {et~al.}(2025{\natexlab{b}}){Wang}, {Tang}, {Wang},
  {Libeskind}, {Tempel}, {Wang}, {Zhang}, {Sheng}, {Yu}, \&
  {Xu}}]{2025ApJ...983..100W}
{Wang}, P., {Tang}, X.-x., {Wang}, H.-d., {et~al.} 2025{\natexlab{b}}, \apj,
  983, 100

\bibitem[{{Wang} {et~al.}(2025{\natexlab{c}}){Wang}, {Wang}, {Rong}, {Wang}, \&
  {Tang}}]{2025JCAP...10..095W}
{Wang}, W., {Wang}, P., {Rong}, Y., {Wang}, H.-d., \& {Tang}, X.-x.
  2025{\natexlab{c}}, \jcap, 2025, 095

\bibitem[{{Welker} {et~al.}(2014){Welker}, {Devriendt}, {Dubois}, {Pichon}, \&
  {Peirani}}]{2014MNRAS.445L..46W}
{Welker}, C., {Devriendt}, J., {Dubois}, Y., {Pichon}, C., \& {Peirani}, S.
  2014, \mnras, 445, L46

\bibitem[{{West} {et~al.}(2017){West}, {de Propris}, {Bremer}, \&
  {Phillipps}}]{2017NatAs...1E.157W}
{West}, M.~J., {de Propris}, R., {Bremer}, M.~N., \& {Phillipps}, S. 2017,
  Nature Astronomy, 1, 0157

\bibitem[{{White}(1984)}]{1984ApJ...286...38W}
{White}, S.~D.~M. 1984, \apj, 286, 38

\bibitem[{{Wittman} {et~al.}(2019){Wittman}, {Foote}, \&
  {Golovich}}]{2019ApJ...874...84W}
{Wittman}, D., {Foote}, D., \& {Golovich}, N. 2019, \apj, 874, 84

\bibitem[{{Yang} {et~al.}(2007){Yang}, {Mo}, {van den Bosch}, {Pasquali}, {Li},
  \& {Barden}}]{2007ApJ...671..153Y}
{Yang}, X., {Mo}, H.~J., {van den Bosch}, F.~C., {et~al.} 2007, \apj, 671, 153

\bibitem[{{Yang} {et~al.}(2012){Yang}, {Mo}, {van den Bosch}, {Zhang}, \&
  {Han}}]{2012ApJ...752...41Y}
{Yang}, X., {Mo}, H.~J., {van den Bosch}, F.~C., {Zhang}, Y., \& {Han}, J.
  2012, \apj, 752, 41

\bibitem[{{Yuan} \& {Wen}(2022)}]{2022MNRAS.516.3159Y}
{Yuan}, Z.~S., \& {Wen}, Z.~L. 2022, \mnras, 516, 3159

\bibitem[{{Zeldovich} {et~al.}(1982){Zeldovich}, {Einasto}, \&
  {Shandarin}}]{1982Natur.300..407Z}
{Zeldovich}, I.~B., {Einasto}, J., \& {Shandarin}, S.~F. 1982, \nat, 300, 407

\bibitem[{{Zel'dovich}(1970)}]{1970A&A.....5...84Z}
{Zel'dovich}, Y.~B. 1970, \aap, 5, 84

\bibitem[{{Zhang} {et~al.}(2015){Zhang}, {Yang}, {Wang}, {Wang}, {Luo}, {Mo},
  \& {van den Bosch}}]{2015ApJ...798...17Z}
{Zhang}, Y., {Yang}, X., {Wang}, H., {et~al.} 2015, \apj, 798, 17

\end{thebibliography}
\bibliographystyle{aasjournal}

\end{CJK*}
\end{document}